\documentclass{article}
\usepackage{bm,latexsym,amsmath,amssymb,amsfonts,fancyhdr,color,graphicx,multirow,slashed,cite,bbold,soul}

\usepackage[a4paper,bottom=3cm,top=2.5cm,head=0mm,width=17cm,dvipdfm]{geometry}
\usepackage[usenames,dvipsnames,svgnames,table]{xcolor}
\usepackage[colorlinks=true,
            linkcolor=blue,
            urlcolor=blue,
            citecolor=green,          
						bookmarks=true,
						bookmarksnumbered=true,
						breaklinks=true,
						pdfstartview=FitBH]{hyperref}
\usepackage[dotinlabels]{titletoc}
\usepackage{titlesec}
\usepackage{authblk,ulem}
\usepackage{multirow}
\usepackage{makecell}
\usepackage{cancel}
\usepackage{enumitem}

\numberwithin{equation}{section}
\allowdisplaybreaks[4]

\titlelabel{\thetitle.\quad \hspace{-0.8em}}
\titlecontents{section}
              [1.5em]
              {\vspace{4mm} \large \bf}
              {\contentslabel{1em}}
              {\hspace*{-1em}}
              {\titlerule*[.5pc]{.}\contentspage}
\titlecontents{subsection}
              [3.5em]
              {\vspace{2mm}}
              {\contentslabel{1.8em}}
              {\hspace*{.3em}}
              {\titlerule*[.5pc]{.}\contentspage}
\titlecontents{subsubsection}
              [5.5em]
              {\vspace{2mm}}
              {\contentslabel{2.5em}}
              {\hspace*{.3em}}
              {\titlerule*[.5pc]{.}\contentspage}

\newcommand{\titledef}{Neutrinoless Double Beta Decay in Light of JUNO First Data}

\newcommand{\mee}{\langle m_{ee}\rangle}
\definecolor{gesfpurple}{rgb}{0.47,0.19,0.42}

\definecolor{gesflanse}{rgb}{0.00,0.50,0.50}

\definecolor{gesfblue}{rgb}{0.08,0.42,0.76}

\definecolor{gesfred}{rgb}{1,0,0}

\definecolor{gesfwhite}{rgb}{1,1,1}

\definecolor{gesfblack}{rgb}{0,0,0}

{}

\newcommand{\dMa}{\delta_{\rm M1}}
\newcommand{\dMb}{\delta_{\rm M2}}
\newcommand{\dMc}{\delta_{\rm M3}}
\newcommand{\dD}{\delta_D}
\newcommand{\Dma}{\Delta m^2_a}
\newcommand{\Dms}{\Delta m^2_s}
\newcommand{\nuless}{0\ensuremath{\nu}2\ensuremath{\beta}}

\newcommand{\vLa}{\overrightarrow{L_1}}
\newcommand{\vLb}{\overrightarrow{L_2}}
\newcommand{\vLc}{\overrightarrow{L_3}}

\newcommand{\ms}{\sum_i m_i}

\newcommand{\Thf}{T_{1/2}^{0\nu}}

\hypersetup{ pdfauthor = {Shao-Feng Ge},
	     pdftitle = {\titledef}, 
	     pdfsubject = {}, 
             pdfkeywords = {}, 
	     pdfcreator = {LaTeX with hyperref package},
	     pdfproducer = {dvips + ps2pdf} }

\usepackage{stackengine}

\newcommand{\gsec}[1]{{\hypersetup{linkcolor=red}Sec.\,\ref{#1}\hypersetup{linkcolor=blue}}}

\newcommand{\geqn}[1]{\hypersetup{linkcolor=blue}Eq.\,(\ref{#1})\hypersetup{linkcolor=blue}}
\newcommand{\gfig}[1]{{\hypersetup{linkcolor=violet}Fig.\,\ref{#1}\hypersetup{linkcolor=blue}}}

\definecolor{Orange}{cmyk}{0,0.61,0.87,0}
\definecolor{JungleGreen}{cmyk}{0.99,0,0.52,0}
\definecolor{OliveGreen}{cmyk}{0.64,0,0.95,0.40}
\definecolor{Brown}{cmyk}{0,0.81,1,0.60}
\definecolor{RoyalBlue}{cmyk}{0.71,0.53,0,0.12}
\definecolor{Gray}{cmyk}{0,0,0,0.40}
\definecolor{LightPink}{cmyk}{0.0,0.25,0,0}
\definecolor{LLightPink}{cmyk}{0.0,0.10,0,0}
\definecolor{LightBlue}{cmyk}{0.25,0,0,0}
\definecolor{LightGray}{cmyk}{0,0,0,0.2}

\definecolor{byzantine}{rgb}{0.74, 0.2, 0.64}
{}

\usepackage{float}

\begin{document}
\fontsize{12pt}{14pt}\selectfont

\title{
       \textbf{\fontsize{15pt}{17pt}\selectfont \titledef}} 
 \author[1,2]{{\large Shao-Feng Ge} \footnote{\href{mailto:gesf@sjtu.edu.cn}{gesf@sjtu.edu.cn}}}
 \affil[1]{State Key Laboratory of Dark Matter Physics, Tsung-Dao Lee Institute \& School of Physics and Astronomy, Shanghai Jiao Tong University, Shanghai 200240, China}
 \affil[2]{Key Laboratory for Particle Astrophysics and Cosmology (MOE) \& Shanghai Key Laboratory for Particle Physics and Cosmology, Shanghai Jiao Tong University, Shanghai 200240, China}
 \author[3]{{\large Chui-Fan Kong} 
 \footnote{\href{mailto:kongcf@ibs.re.kr}{kongcf@ibs.re.kr}}}
 \affil[3]{Particle Theory and Cosmology Group (PTC), Center for Theoretical Physics of the Universe (CTPU), Institute for Basic Science, Daejeon 34126, Republic of Korea}
 \author[4]{{\large Manfred Lindner} \footnote{\href{mailto:lindner@mpi-hd.mpg.de}{lindner@mpi-hd.mpg.de}}}
 \affil[4]{Max-Planck-Institut f\"{u}r Kernphysik, Heidelberg 69117, Germany}
 \author[1,2]{{\large Jo\~ao Paulo Pinheiro} \footnote{\href{mailto:joaopaulo.pinheiro@fqa.ub.edu}{joaopaulo.pinheiro@fqa.ub.edu}}}

\date{\today}

\maketitle

\begin{abstract}
\fontsize{12pt}{14pt}\selectfont
The first results from the JUNO reactor neutrino oscillation
experiment improve our knowledge of neutrino masses and
mixing parameters, especially the solar angle $\theta_s \equiv \theta_{12}$
and the solar mass squared difference $\Delta m^2_s \equiv \Delta m^2_{21}$.
We discuss the implications of these
results on neutrinoless double beta decay by itself and
in combination with the global fit of neutrino oscillation
experiments, the JUNO first data, and cosmological constraints
on the neutrino mass sum.
For the effective mass $\langle m_{ee}\rangle$, the uncertainties in
its lower limits for both mass orderings and upper limits
for the normal ordering are largely reduced. Since the cosmological
CMB and DESI BAO data put a stringent
constraint on the neutrino
mass scale, we also show how the probability distribution
of both the real and imaginary parts of the effective mass
$\langle m_{ee}\rangle$ on the complex plane is affected.
Especially, the funnel region with
$|\langle m_{ee}\rangle| \lesssim 1$\,meV receives larger chance to happen.
Correspondingly, the chance of determining the two Majorana
CP phases simultaneously in this region also increases with
reduced uncertainty.
\end{abstract}


\newpage

\section{Introduction}

The baryon asymmetry of the Universe, namely that almost 
no anti-matter exists in Nature \cite{Cohen98,Canetti12},
is one of the main problems in contemporary astroparticle 
physics and cosmology. An inflationary phase at the beginning 
of our Universe dilutes all pre-existing matter \cite{Guth81}
which requires a mechanism to understand the observed baryon 
asymmetry. The Sakharov conditions \cite{Sakharov:1967dj} 
summarize the essential ingredients which must be fulfilled. 
One very well motivated mechanism is leptogenesis \cite{Fukugita86} 
which is typically related to 
Majorana neutrino masses with additional CP phases emerging by the seesaw
mechanism \cite{Fritzsch:1975sr,Minkowski:1977sc,Yanagida:1979as,Gell-Mann:1979vob,Glashow:1979nm,Mohapatra:1979ia,Konetschny:1977bn,Magg:1980ut,Schechter:1981cv,Cheng:1980qt,Lazarides:1980nt,Mohapatra:1980yp,Foot:1988aq,Ma:1998dn}.

Verifying the Majorana nature of neutrinos is therefore 
very important for justifying the vanilla leptogenesis mechanism.
Although the ultra-violet physics at high energy is difficult
to observe with current technology, with two exceptions of
cosmic string \cite{Dror:2019syi} and cosmological collider
\cite{Cui:2021iie}, the  nature
inherited by light active neutrinos can manifest itself
in the neutrinoless double beta decay ($0 \nu 2 \beta$) process
where a mother nuclei decays into a daughter nuclei plus
two electrons, ${}^Z_A N \rightarrow {}^{Z+2}_A N + 2 e^-$
\cite{Dolinski:2019nrj,Shergold:2021evs,Agostini:2022zub,Cirigliano:2022oqy}.
With two electrons appearing in
the final state, such process indicates that the lepton
number is violated by two. In the minimal scenario,
the $0 \nu 2 \beta$ decay can be induced by an effective Majorana
mass $m_{ee}$ of the light active neutrinos in the electron
flavor. Reciprocally, a lepton number violating process
would finally contribute to a light 
Majorana mass \cite{Schechter82}.

However, the $0 \nu 2 \beta$ decay process involves not just
particle physics, but also nuclear reaction and atomic effects.
The decay half-life can be expressed as
\begin{equation}
  \left( T^{0\nu}_{1/2} \right)^{-1}
\equiv
  G^{0\nu} |M^{0\nu}|^2 |\mee|^2,
\label{eq:Thalf}
\end{equation}
where $G^{0\nu}$ is the phase space factor
\cite{Kotila:2012zza,Kotila:2013gea,Mirea:2014dza,Neacsu:2015uja}
associated with the decay kinematics and atomic
physics, $M^{0\nu}$ is the nuclear matrix element (NME),
and $\mee$ is the effective electron neutrino mass
from particle physics.
Of these three parts, the phase space factor can be calculated very
precisely while the NME suffers a lot from theoretical
uncertainties \cite{Agostini:2022zub}.
The NME calculations involve complex nuclear
many-body physics and depend on the nuclear structure models. 
There are multiple theoretical approaches such as the quasiparticle
random phase approximation (QRPA) \cite{qrpa1,qrpa2,qrpa3,qrpa4,qrpa5},
interacting shell mode (ISM) \cite{shell1,shell2,shell3,shell4},
energy density functional methods (EDF) \cite{edf1,edf2,edf3},
interacting boson model (IBM) \cite{ibm1,ibm2}, and ab initio
techniques \cite{abinitio1,abinitio2,abinitio3,abinitio4,abinitio5,abinitio6,abinitio7}.
Their predictions typically differ by a factor of $2 \sim 3$,
representing the main source of systematic uncertainty from
the nuclear physics side \cite{Agostini:2022zub}. 

On the other hand, the particle physics side has improved
quite significantly. The improvement comes from neutrino
oscillation experiments, beta decay spectrum measurement
around the end point, and cosmology. Especially, there
is complementarity between the reactor neutrino oscillation
experiments and $0 \nu 2 \beta$ decay
\cite{Pascoli:2005zb, Choubey:2005rq,
Lindner:2005kr,Dueck:2011hu,Xing:2015zha, GE1, GE2,Cao:2019hli,Ge:2019ldu,Huang:2020mkz,Denton:2023hkx}
since both involves only the electron flavor and share the
same set of neutrino oscillation parameters. To be exact,
only two mixing angles, the reactor angle
$\theta_r \equiv \theta_{13}$ and the
solar angle $\theta_a \equiv \theta_{12}$ can appear
in the effective Majorana mass $m_{ee}$ for
$0 \nu 2 \beta$ decay and the disappearence
probability $P_{ee}$ of reactor neutrino oscillation.
In addition, the solar mass squared difference
$\Delta m^2_s \equiv \Delta m^2_{21}$ and
the atmospheric one
$\Delta m^2_a \equiv \Delta m^2_{31}$ can also enter.
It is interesting to see that these four
oscillation parameters can be precisely measured
with the combination of Daya Bay and JUNO
\cite{Ge:2012wj,JUNO:2015zny}. Especially, the recent global
fits \cite{NuFIT6} and JUNO first data \cite{JUNO1st} have
significantly reduced the uncertainty of the solar
angle $\theta_s$. In addition, the KATRIN experiment
provides the best measurement of the beta decay spectrum
to put a $\sum_i m_i \leq 0.45$\,eV limit at 90\%\,C.L.
(confidence level) \cite{KATRIN2022,KATRIN2024}.
The most stringent bound on the absolute neutrino mass
comes from the recent DESI BAO data \cite{DESI2024,DESI2024b},
plus other
existing observations such as CMB \cite{Planck:2018vyg}.

Concerning the recent progress in neutrino experiments
and cosmology, we would like to show how
these affect the prospects of the $\nuless$ decay.
We first summarize the particle physics contribution to
the $0 \nu 2 \beta$ decay in \gsec{sec:theory0vbb}.
The following \gsec{sec:nulessJUNO} studies the influence
of the recent global fit and JUNO first data release in
particular on the effective mass and half-life time.
We further study the possibility for the effective
mass $\mee$ to fall into the funnel region and
the prospect of determining the two Majorana CP phases
simultaneously in \gsec{sec:majo_triang}. The conclusion
of this paper can be found in \gsec{sec:conclusion}.

\section{Effective Majorana Mass for $\texorpdfstring{\nuless}{0 nu 2 beta}$ Decay}
\label{sec:theory0vbb}

As summarized in \geqn{eq:Thalf}, the half-life time
$T^{0 \nu}_{1/2}$ contains three parts from atomic
physics, particle physics, and nuclear physics, respectively.
This section summarizes the effective mass $\mee$
from the particle physics side. With $\mathcal O(1)$\,MeV
energy release, the $\nuless$ process can only produce
two electrons which explain why only $\mee$ for the
electron flavor can be probed.


The effective mass $\mee$ is defined as:
\begin{equation}
\mee \equiv \sum_i m_i U_{ei}^2,
\label{eq:mee_definition}
\end{equation}
where $m_i$ are the neutrino mass eigenvalues and $U_{ei}$
are the PMNS mixing matrix elements.
We follow the PMNS matrix parametrization \cite{PDG24},
\begin{equation}
  U
\equiv
  \mathcal{P}
\left\lgroup
\begin{array}{ccc}
  c_s c_r & s_sc_r & s_re^{-i\dD} \\
- c_a s_s - s_a s_r c_se^{i\dD} & c_ac_s - s_as_rs_se^{i\dD} & s_ac_r \\
  s_a s_s - c_as_rc_se^{i\dD} & -s_ac_s - c_as_rs_se^{i\dD} & c_ac_r
\end{array}
\right\rgroup
\mathcal{Q},
\label{eq:U}
\end{equation}
with $\mathcal{P} \equiv \text{diag}\{e^{-i\beta_1}, e^{-i\beta_2}, e^{-i\beta_3}\}$
and $\mathcal{Q} \equiv \text{diag}\{e^{-i\dMa/2}, e^{-i\dMb/2}, e^{-i(\dMc - \dD)/2}\}$.
The three phases $\beta_i$ in $\mathcal{P}$ are unphysical while 
$\mathcal{Q}$ contains two physical Majorana phases.
We set $\dMb = 0$ but take $\dMa$ and $\dMc$ as the two
independent physical Majorana phases. The benefit of using such
parametrization is that when the lightest mass vanishes, $m_1 = 0$ for
the normal ordering (NO) case or $m_3 = 0$ for the inverted ordering (IO),
one physical Majorana CP phase would automatically disappear altogether.
For convenience, we labeled the three mixing angles according
to the major type of experiments that fixes their values,
$\theta_a \equiv \theta_{23}$ for the atmospheric angle,
$\theta_r \equiv \theta_{13}$ for the reactor angle, and
$\theta_s \equiv \theta_{12}$ for the solar angle. 

Note that the third element of $\mathcal{Q}$ is defined as
$e^{-i(\dMc - \dD)/2}$ where the Dirac CP phase $\delta_D$
is to cancel the same Dirac CP phase associated with the
reactor angle $s_r$ of $U$ as defined in \geqn{eq:U},
\begin{equation}
  \mee
=
  m_1 |U_{e1}|^2 e^{i\dMa}
+ m_2 |U_{e2}|^2
+ m_3 |U_{e3}|^2 e^{i\dMc}.
\label{eq:mee_formula0}
\end{equation}
A particularly illuminating perspective emerges when
treating $\mee$ as a vector sum in the complex plane
which can give a geometrical view
\cite{Vissani:1999tu,Xing:2014yka,GE2}.
We can decompose it into three contributions,
\begin{equation}
  \mee
\equiv
  \vLa + \vLb + \vLc,
\label{eq:triangle}
\end{equation}
with each vector being defined as,
\begin{eqnarray}    
\vLa \equiv m_1 |U_{e1}|^2 e^{i\dMa},
\qquad
\vLb \equiv m_2 |U_{e2}|^2,
\qquad
\vLc \equiv m_3 |U_{e3}|^2 e^{i\dMc}.
\end{eqnarray}
With our phase convention $\dMb = 0$, $\vLb$ lies
along the real axis while $\vLa$ and $\vLc$ rotate
with the corresponding Majorana CP phases $\dMa$
and $\dMc$, respectively. This geometric
representation will be discussed in detail in
\gsec{sec:majo_triang}, where we explore how
the interplay between these vectors constrains
the allowed values of $\mee$.

Using the specific parametrization of the PMNS
matrix as defined in \geqn{eq:U}, we can 
write the mixing matrix elements explicitly,
\begin{equation}
  \mee
=
  c^2_s c^2_r m_1 e^{i\dMa}
+ s^2_s c^2_r m_2
+ s^2_r m_3 e^{i\dMc} ,
\label{eq:mee_formula}
\end{equation}
where $(c_\alpha, s_\alpha) \equiv (\cos\theta_\alpha, \sin\theta_\alpha)$.
Note that the atmospheric mixing angle $\theta_a$
does not appear explicitly in $\mee$, which avoids
any ambiguity related to the determination of the
octant of $\theta_{a}$. The $\nuless$ decay shares
four oscillation parameters with the reactor neutrino
oscillation. Besides the reactor angle $\theta_r$ and
the solar angle $\theta_s$, both the solar mass squared
difference $\Dms \equiv m^2_2 - m^2_1$ and the
atmospheric mass squared difference
$\Dma \equiv |\Delta m^2_{3\ell}| = |m^2_3 - m^2_\ell|$
where $\ell=1$ for NO ($m_1 < m_2 < m_3$) and
$\ell=2$ for IO ($m_3 < m_1 < m_2$)
also appear. To make the physics behind the mass
squared differences more explicit, we have also
adopted their labels according to the major type
of observations from which their values are determined.


The behavior of $\mee$ significantly depends on the
ordering of the neutrino mass, with fundamentally
different phenomenological implications for each scenario. 
The main distinction arises from the lightest mass,
$m_1$ for NO and $m_3$ for IO. This difference has direct
consequences for the allowed range of $\mee$,
particularly in the limit where the lightest mass
is smaller than $0.1$\,eV. Note that distinguishing
the mass ordering is the major physics goal of the
JUNO experiment \cite{JUNO:2015zny}.

For NO, we express the heavier masses ($m_2$ and $m_3$)
in terms of $m_1$ and the measured mass-squared differences,
\begin{equation}
  \mee
=
  m_1 c^2_s c^2_r e^{i\dMa}
+ \sqrt{\Dms + m_1^2} c^2_r s^2_s
+ \sqrt{\Dma +m_1^2} s^2_r e^{i\dMc}.
\label{eq:mee_NO}
\end{equation}
In this case, $\mee$ can vanish if the Majorana phases conspire to produce a complete destructive interference between the remaining contributions. Thus, observing $\mee = 0$ would not exclude the Majorana nature of neutrinos under NO.

For IO, the masses are parametrized in terms of the lightest mass $m_3$:
\begin{equation}
  \mee
=
  \sqrt{m_3^2 + \Dma - \Dms} c^2_s c^2_r e^{i\dMa}
+ \sqrt{m_3^2 + \Dma} c^2_r s^2_s
+ m_3 s^2_r e^{i\dMc}.
\label{eq:mee_IO}
\end{equation}
The main feature of IO is that $\mee$ possesses a non-zero lower bound. This lower bound, which depends on the mixing angles and mass splitting, provides a testable prediction that can potentially discriminate between mass orderings for Majorana neutrinos. The minimum values allowed for $\mee$ for IO will be explored in the next \gsec{sec:nulessJUNO}.

\section{Status of \texorpdfstring{\nuless}{0nu2beta} Observables after JUNO Results}
\label{sec:nulessJUNO}

The interplay between the neutrino oscillation parameters
and the predictions for neutrinoless double-beta decay has
been extensively studied in the literature, with detailed
analysis of how experimental uncertainties propagate into
the allowed ranges for the effective Majorana mass $\mee$
\cite{Pascoli:2002xq,Pascoli:2005zb,Choubey:2005rq,Lindner:2005kr,Dueck:2011hu, GE1, Denton:2023hkx}.
Especially, the uncertainty in the solar angle $\theta_s$
has significant effect on the upper and lower limits of
the effective mass $\mee$ \cite{Dueck:2011hu, GE1}.
In this section, we perform an updated analysis of \cite{GE1}
by incorporating the latest global fit results from neutrino
oscillation experiments \cite{NuFIT6} and more crucially
the first measurement of solar oscillation parameters
$\sin^2\theta_s$ and $\Delta m^2_s$
from the JUNO experiment \cite{JUNO1st}.
Our goal is to quantify how the improved determination
of the solar mixing angle $\theta_s$ from the JUNO initial
59.1-day dataset impacts the theoretical predictions
of $\mee$.

As commented in the previous \gsec{sec:theory0vbb},
different neutrino mass orderings have very different
phenomenological consequences on the effective $\mee$
of $\nuless$ decay. While $\mee$ could vanish for NO,
it is bounded from not just above but also below for IO,
\begin{subequations}
\begin{align}
  \mee^{\rm IO}_{\rm max}
& \equiv
  \sqrt{m^2_3 + \Dma} c^2_s c^2_r
+ \sqrt{m^2_3 + \Dma - \Dms} s^2_s c^2_r
+ m_3 s^2_r \,,
\\
  \mee^{\rm IO}_{\rm min}
& \equiv
  \sqrt{m^2_3 + \Dma} c^2_s c^2_r
- \sqrt{m^2_3 + \Dma - \Dms} s^2_s c^2_r
- m_3 s^2_r \,.
\end{align}
\label{eq:IH-limits}
\end{subequations}
The upper and lower limits happen when the two
Majorana phases take specific values to give constructive
or destructive interference, respectively.

\begin{figure}[t]
\centering
\includegraphics[width=0.32\linewidth]{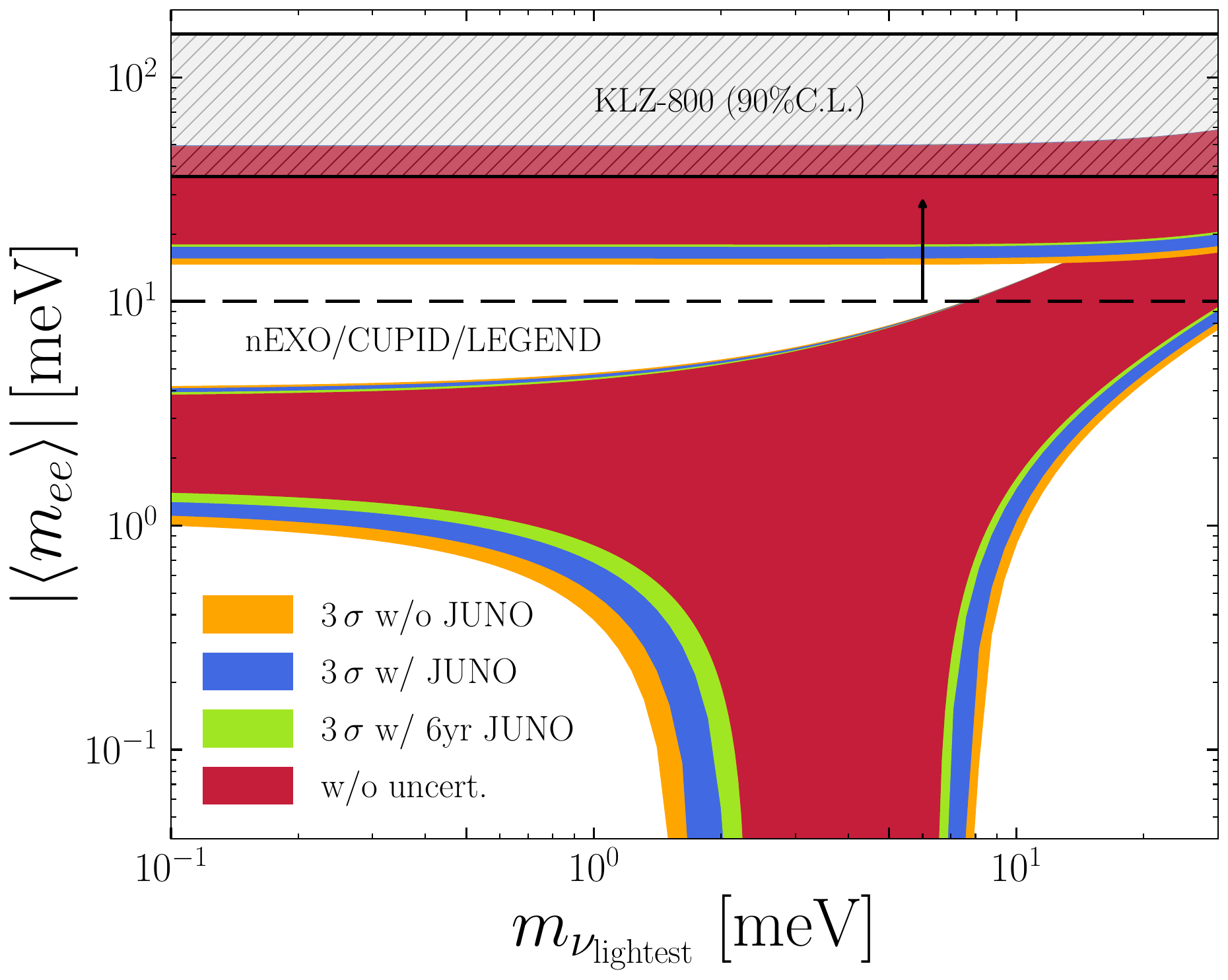}
\includegraphics[width=0.32\linewidth]{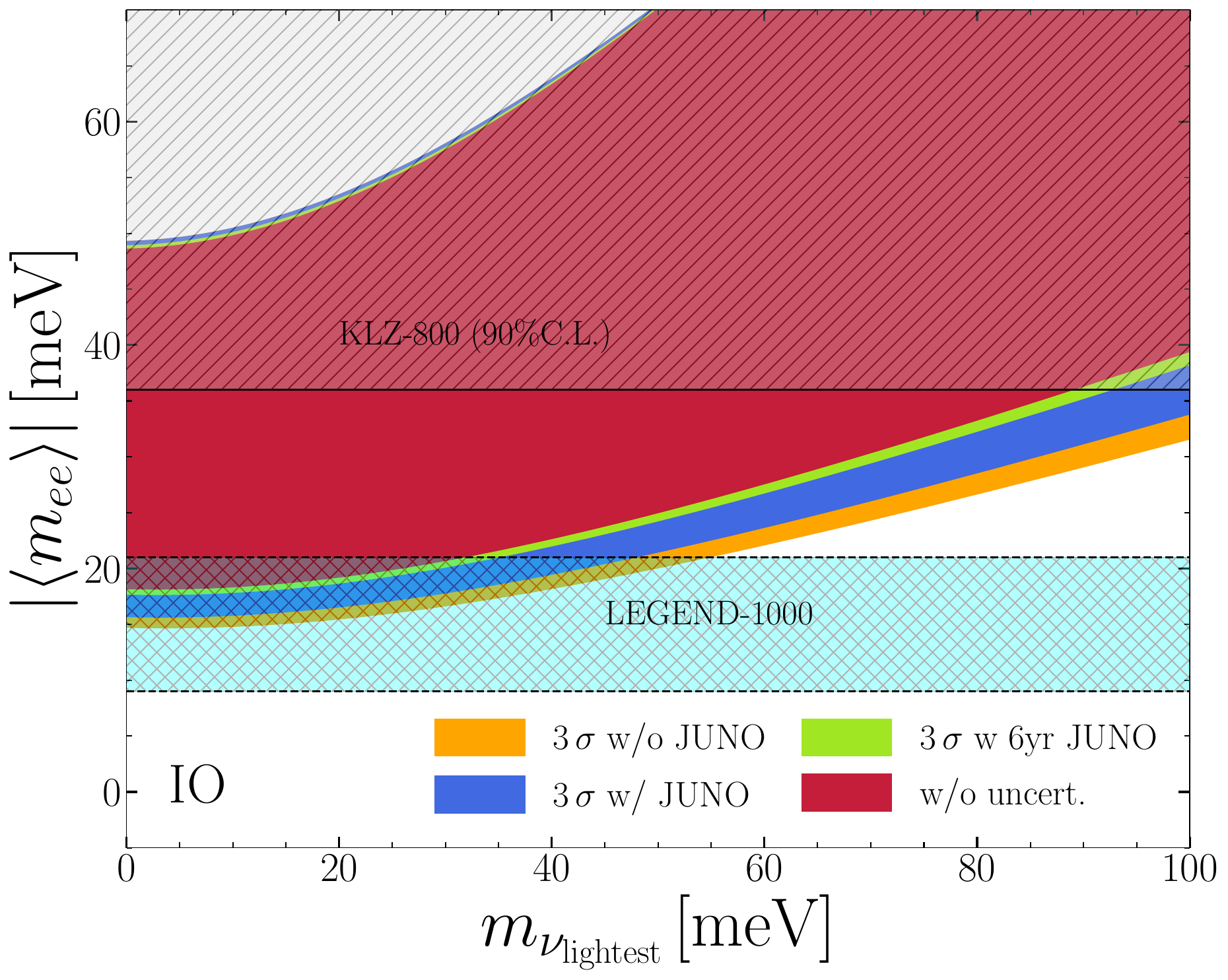}
\includegraphics[width=0.32\linewidth]{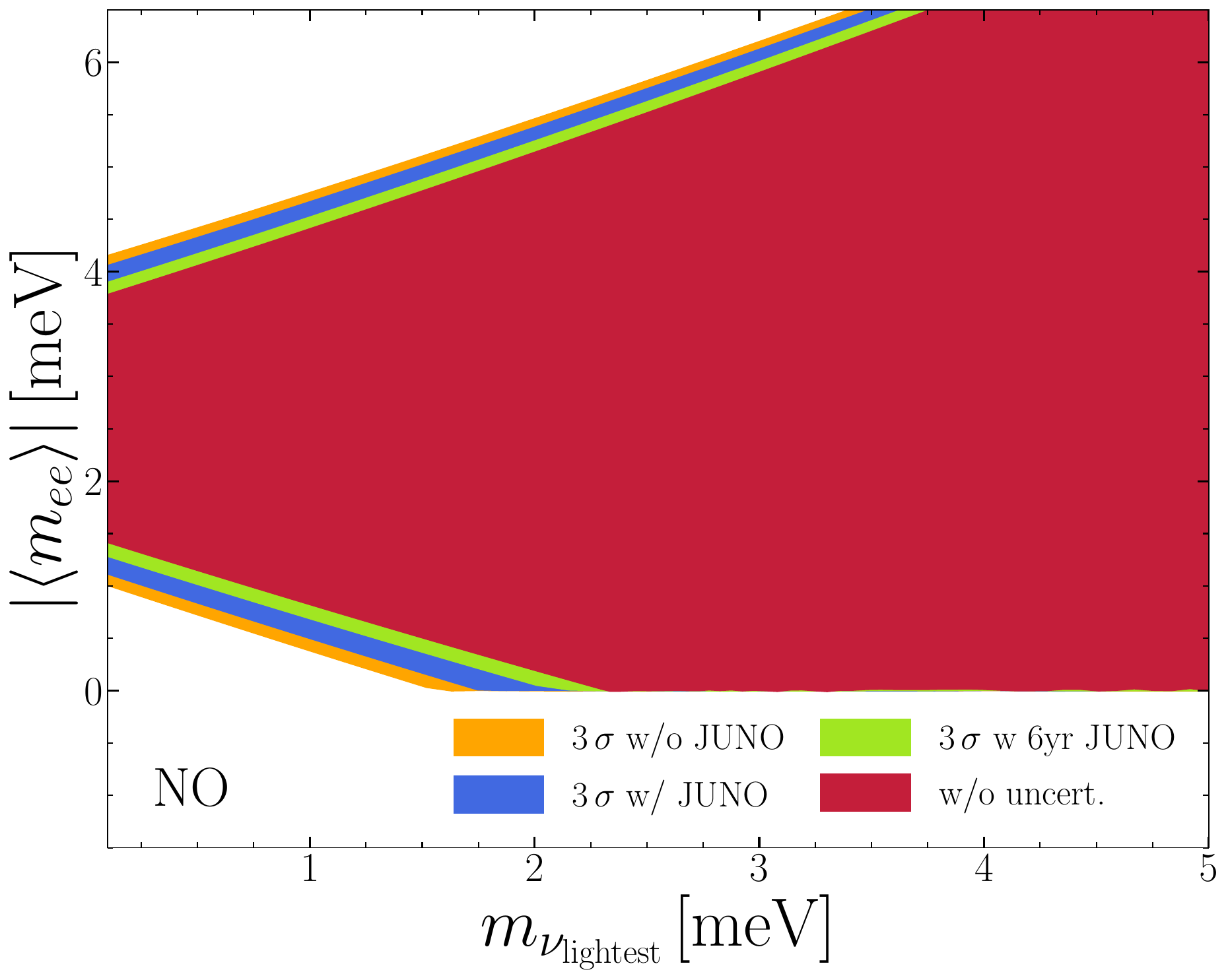}
\caption{The particle physics uncertainties for the
$0 \nu 2 \beta$ decay effective mass $\langle m_{ee} \rangle$.
For comparison and easy viewing, three panels are shown:
{\bf (Left)} overall features for both NO and IO;
{\bf (Middle)} zoomed-in linear plot for IO;
{\bf (Right)} zoomed-in linear plot of the low-mass horizontal
branch for NO. The allowed regions of the effective mass
$\langle m_{ee} \rangle$ have been demonstrated with four
different uncertainty configurations:
{\bf (1)} the current $3\,\sigma$ range from NuFIT 6.0 global fit (orange);
{\bf (2)} the updated $3\,\sigma$ range with the JUNO first data release (blue);
{\bf (3)} the expected $3\,\sigma$ range with full JUNO (green);
{\bf (4)} the ultimate scenario with negligible uncertainty (red).
The current best limit on the effective mass $\langle m_{ee} \rangle$
from KamLAND-Zen with 800\,kg Xenon mass has
been added as hashed region with solid boundaries
while the expected sensitivities from CUPID,
LEGEND1000 and nEXO are shown in the left plot as dashed line without uncertainties of nuclear matrix elements. The middel plot includes the sensitivity of LEGEND1000 including uncertainties of nuclear matrix elements.}
\label{fig:0vbb_comparison}
\end{figure}

As shown in the left panel of \gfig{fig:0vbb_comparison},
these two limits become flat when the lightest mass,
$m_1$ for NO and $m_3$ for IO, approaches zero.
More generally, a vanishing lightest mass would not
have much effect on the $\nuless$ effective mass $\mee$.
Then, only the two mixing angles ($\theta_r$ and $\theta_a$)
and the two mass squared differences ($\Delta m^2_a$
and $\Delta m^2_s$) enters. For IO,
\begin{eqnarray}
  \mee^{\rm IO}_{\rm max}
\approx
  \sqrt{\Dma} c_r^2,
\qquad
  \mee^{\rm IO}_{\rm min}
\approx
  \sqrt{\Dma} (c_s^2-s_s^2)c_r^2.
\label{eq:IH-limit_min}
\end{eqnarray}
Although both mass squared differences are involved
in the electron flavor effective mass $\mee$, the
limits are dominated by the atmohsperic one $\Dma$ with
much larger value. Further, the upper limit
$\mee^{\rm IO}_{\rm max}$ even becomes independent
of the solar angle $\theta_s$. Since both the
atmospheric mass squared difference $\Dma$ and the reactor
angle $\theta_r$ have been measured quite precisely
with just roughly 1\% error \cite{NuFIT6},
there is almost no uncertainty in the upper limit for IO.
This has been clearly shown in \gfig{fig:0vbb_comparison}
as the filled regions for IO have roughly the same
upper boundary.

For comparison, the lower limit $\mee^{\rm IO}_{\rm min}$
depends on not just $\Dma$ and the reactor angle
$\theta_r$ but also the
solar angle $\theta_s$. Although $\theta_s$ has also
been measured very precisely with only 2\% uncertainty
\cite{NuFIT6}, the combination $\cos 2 \theta_s$ that
appears in $\mee^{\rm IO}_{\rm min}$ can vary by
1/3 within the $3\,\sigma$ range. Even in the left panel
of \gfig{fig:0vbb_comparison} with logarithmic scale,
the effect of improving the solar angle measurement with
JUNO can be clearly seen. Since the lower
limit $\mee^{\rm IO}_{\rm min}$ corresponds to the
maximal value of half-life time $T^{0\nu}_{1/2}$
according to \geqn{eq:Thalf}, such uncertainty would
have significant consequences when interpreting the
experimental data. Any improvement in the determination
of $\theta_s$ would significantly reduce the uncertainty
on the lower bound of $\mee$ for IO \cite{Pascoli:2002xq,GE1}.

For NO, the effective mass in \geqn{eq:mee_NO} is also
bound from below and above with a vanishing lightest
mass $m_1 \rightarrow 0$,
\begin{align}
  \mee^{\rm NO}_{\rm min}
\approx
  \sqrt{\Delta m^2_s} c^2_r s^2_s
- \sqrt{\Delta m^2_a} s^2_r,
\qquad
  \mee^{\rm NO}_{\rm max}
\approx
  \sqrt{\Delta m^2_s} c^2_r s^2_s
+ \sqrt{\Delta m^2_a} s^2_r.
\label{eq:limitsNO}
\end{align}
These two limits contain two mass squared differences
and two mixing angles. Since the reactor angle
$\theta_r$ is not large, its consine is roughly
$c_r \approx 1$. The relative size between the
two terms in \geqn{eq:limitsNO} is then determined
by comparing the ratios $\Dms/\Dma$ and $s^2_r / s^2_s$.
According to the global fit \cite{NuFIT6},
$\sqrt{\Dms / \Dma} \approx 0.17$ and
$s^2_r / s^2_s \approx 0.07$ which means the
first term $\sqrt{\Dms} c^2_r s^2_s$ dominates.
Note that the solar angle $s^2_s$ appears in the
first term. The current $3\,\sigma$ variation of $s^2_s$
can be as large as 23\% \cite{NuFIT6}.
Although the solar mass squared difference $\Dms$ 
itself varies by $\sim 15\%$ at $3\,\sigma$, its 
contribution to $\mee$ is only $\sim 7.5\%$.
So $\theta_s$ is the dominant source of uncertainty for NO. 
With all factors combined, the total effect of those
uncertainties from the current data can be clearly
seen in the orange bands of the left and right panels
of \gfig{fig:0vbb_comparison}. 
The difference is much more visible for the lower
limits. This is because the variation in
the first term $\sqrt{\Dms} c^2_r s^2_s$
actually affects the central values as well
and the whole NO branch in the limit of
vanishing lightest mass moves vertically.
While the lower limit $\mee^{\rm NO}_{\rm min}$
is different between the two terms in \geqn{eq:limitsNO},
the upper one $\mee^{\rm NO}_{\rm max}$ is a sum
with larger value. Then the relative variation
is much smaller for the upper limit.

To make the influence of $\theta_s$ on the effective
mass $\mee$ more visible, we have included two linear
scale plots for IO and NO as the middle and right
panels of \gfig{fig:0vbb_comparison}. 
From this figure, it is also possible to 
infer from the orange bands both the minimum and 
maximum values of the effective mass 
for each mass ordering in the vanishing lightest mass scenario.
Explicitly, their $3\,\sigma$ intervals are 
$\mee^{\rm IO}_{\rm min}=(15.72\sim22.28)\,\text{meV}$, 
$\mee^{\rm NO}_{\rm min}=(1.188\sim1.828)\,\text{meV}$, and 
$\mee^{\rm NO}_{\rm max}=(3.404\sim4.028)\,\text{meV}$, 
resulting from the propagation of the oscillation parameter uncertainties.
While these uncertainties appear modest in $\mee$,
their influence on the $\nuless$
experiments is much more significant \cite{GE1}.
Especially, the half-life
time scales inversely with the effective mass squared,
$T^{0\nu}_{1/2} \propto 1 / |\mee|^2$ according to
\geqn{eq:Thalf}. The variation in $\mee$ would get
amplified when propagating to the half-life time.
Further, to reach the required sensitivity on
$T^{0\nu}_{1/2}$ the expected experimental exposure
scales \cite{Avignone:2005cs},
$M \times t \propto (T^{0\nu}_{1/2})^2$ where
$M$ is the detector target mass and $t$ is the running
time, with the half-life time squared. Finally,
the experimental exposure can differ by a factor
of 4.3 for IO and 2.5 for NO. The improvement on
the uncertainty from the particle physics side by
JUNO is expected to significantly reduce the uncertainty
in the experimental design and data interpretation.

Following the first data release from JUNO, which
reported 59.1 days of data taking \cite{JUNO1st},
the precision on the solar
angle $\theta_s$ and the solar mass splitting $\Dms$ has
been significantly improved to
$\sin^2\theta_{s} = 0.3092\pm 0.0087$ and
$\Dms = (7.50\pm 0.12)\times 10^{-5}\,$eV$^2$.
This corresponds to a reduction of uncertainty by
approximately a factor of $0.741$ at $3\,\sigma$ than the
global fit results from NuFIT 6.0 \cite{NuFIT6}.
According to \geqn{eq:IH-limit_min}, this translates
to around 22\% reduction in the uncertainty of
$\mee^{\rm IO}_{\rm min} \propto c^2_s - s^2_s$
at $3\,\sigma$. This improvement was achieved
through JUNO's unique capability to observe the
low-frequency mode of the reactor antineutrino
oscillation \cite{Ge:2012wj},
\begin{align}
  P_{ee}
\approx
  1
- c^4_r (c^2_s - s^2_s)
  \sin^2 \left( \frac {\Dms L}{4 E_\nu} \right).
\end{align}
For simplicity, we have omitted those high-frequency
mode modulated by $\Dma$ which are important for the
mass ordering measurement. The amplitude of the
low-frequency mode is approximately $c^2_s - s^2_s$
which is exactly the one that appears in the lower
limit $\mee^{\rm IO}_{\rm min}$ for IO. With large
statistics at JUNO, the solar angle can be measured
with very high precision. At the same time, the
mode frequency or equivalently the solar mass squared
difference $\Dms$ can also be measured with comparable
precision. This impressive result
within just 59.1 days of data taking demonstrates
JUNO's potential to achieve sub-percent precision
on solar parameters with the full dataset, making
it the leading experiment for solar mixing angle
determination in the coming years.

The improved determination of $\theta_s$ and $\Dms$ directly
translates into enhanced constraints on the effective
Majorana mass $\mee$, as shown in \gfig{fig:0vbb_comparison}.
It narrows the allowed band for $\mee$ in both mass
orderings. Particularly, the lower bound $\mee^{\rm IO}_{\rm min}$
is most relevant for the sensitivity goals of next-generation
$\nuless$ experiments.

For comparison, \gfig{fig:0vbb_comparison} shows
four uncertainty scenarios. First, the theoretical
prediction with perfect knowledge of oscillation
parameters (red) that represents the ultimate goal
of mitigating the uncertainty from the particle
physics side to a negligible extend \cite{GE1}.
Second, the current $3\,\sigma$ range with the
pre-JUNO global fits (orange) that shows what other
experiments have achieved altogether \cite{NuFIT6}.
Between these two cases, there is a huge space
which is exactly the existing uncertainty from
particle physics. In the middle, we have show two
uncertainty scenario from the JUNO experiment.
The blue band shows the reduced uncertainty
with the JUNO first data release \cite{JUNO1st}.
The comparison between the orange (pre-JUNO) and
blue (post-JUNO) bands clearly illustrates the
impact of the improved solar angle measurement
by JUNO with a noticeable reduction in the
width of the allowed regions for both mass orderings.
The green one corresponds to the expected
sensitivity with 6 years of data collection
\cite{JUNO:2015zny}.

From \gfig{fig:0vbb_comparison}, it is possible to observe 
that the JUNO determination of $\theta_{s}$ sharpens the lower edge of the IO band (compare blue vs. orange), with total uncertainty in $\mee^{\rm IO}_{\rm min}$ at the $3\sigma$ C.L. by:
\begin{eqnarray}
  \mee^{\rm IO}_{\rm min}
=
  (16.36 \sim 21.48)\,\mathrm{meV}.
\end{eqnarray}
For comparison, we have also shown those expected
sensitivities from $\nuless$ decay experiments as
horizontal lines. The most stringent current bounds
come from KamLAND-Zen ($^{136}$Xe), yielding
$\Thf > 2.3 \times 10^{26}$\,yr at $90\%$ C.L.,
which translates to $\mee < (36 \sim 156)$\,meV
depending on the nuclear matrix elements (NME)
\cite{KamLAND-Zen:2024eml,KamLAND-Zen:2022tow}.
This bound is indicated as the hashed region with
solid boundaries in the left and middle panels of
\gfig{fig:0vbb_comparison}. The current experiments
are beginning to probe the IO parameter space,
particularly in the region where
$m_{\text{lightest}} \gtrsim 10$\,meV.  
The next-generation $0\nu\beta\beta$ experiments,
including nEXO \cite{nEXO1,nEXO2}, LEGEND-1000
\cite{LEGEND,LEGEND:2021bnm}, and CUPID
\cite{CUPID1,CUPID2}, are projected to reach
sensitivities of $\mee < (5 \sim 12)$\,meV, which
would fully cover the IO prediction band and
begin to probe the NO region. 

\begin{figure}[t]
\centering
\includegraphics[width=0.32\linewidth]{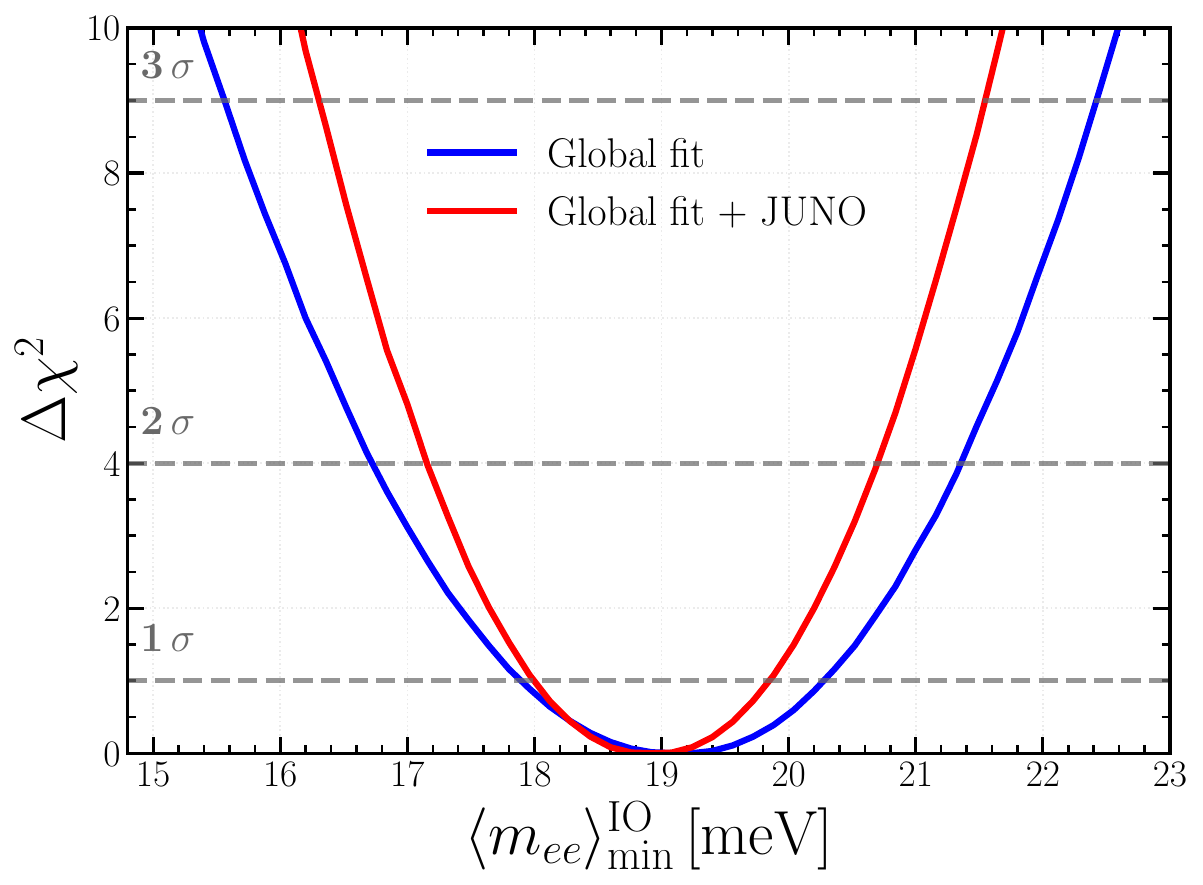}
\includegraphics[width=0.32\linewidth]{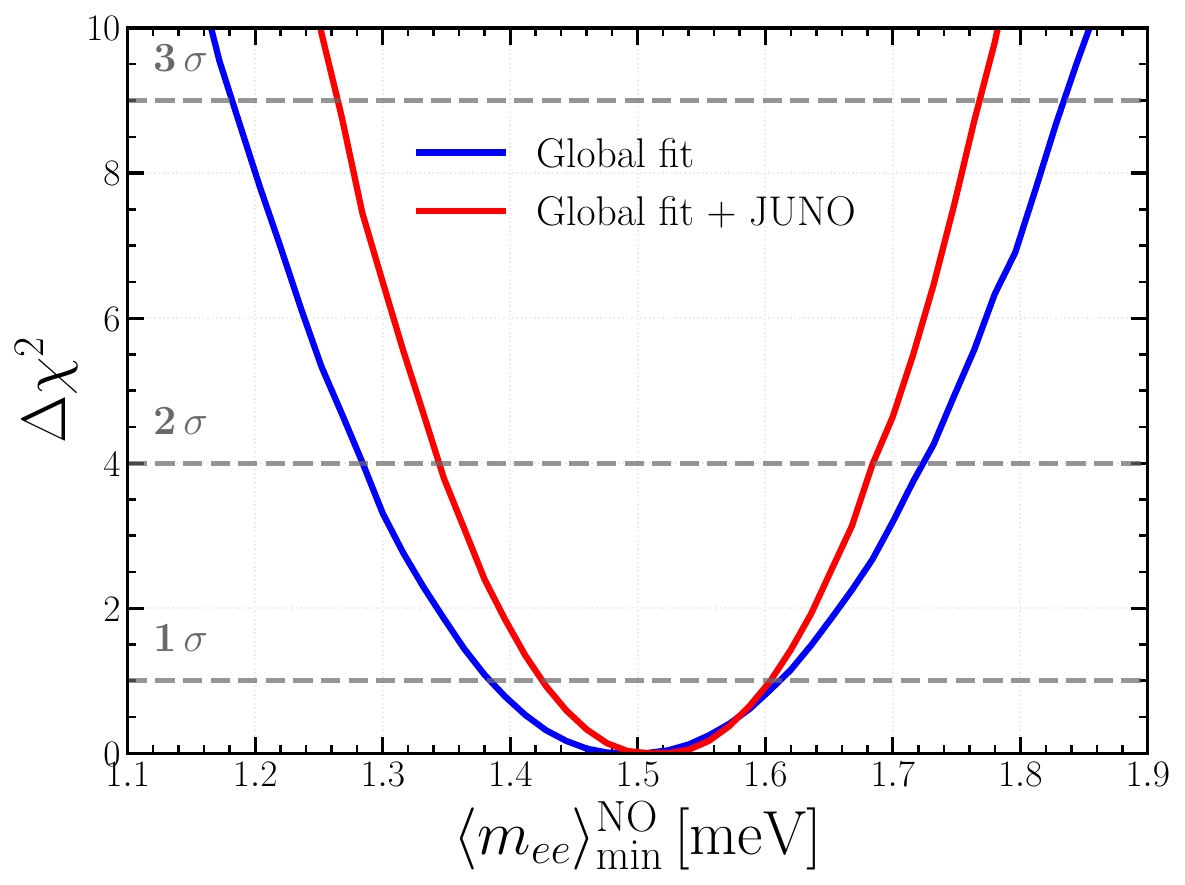}
\includegraphics[width=0.32\linewidth]{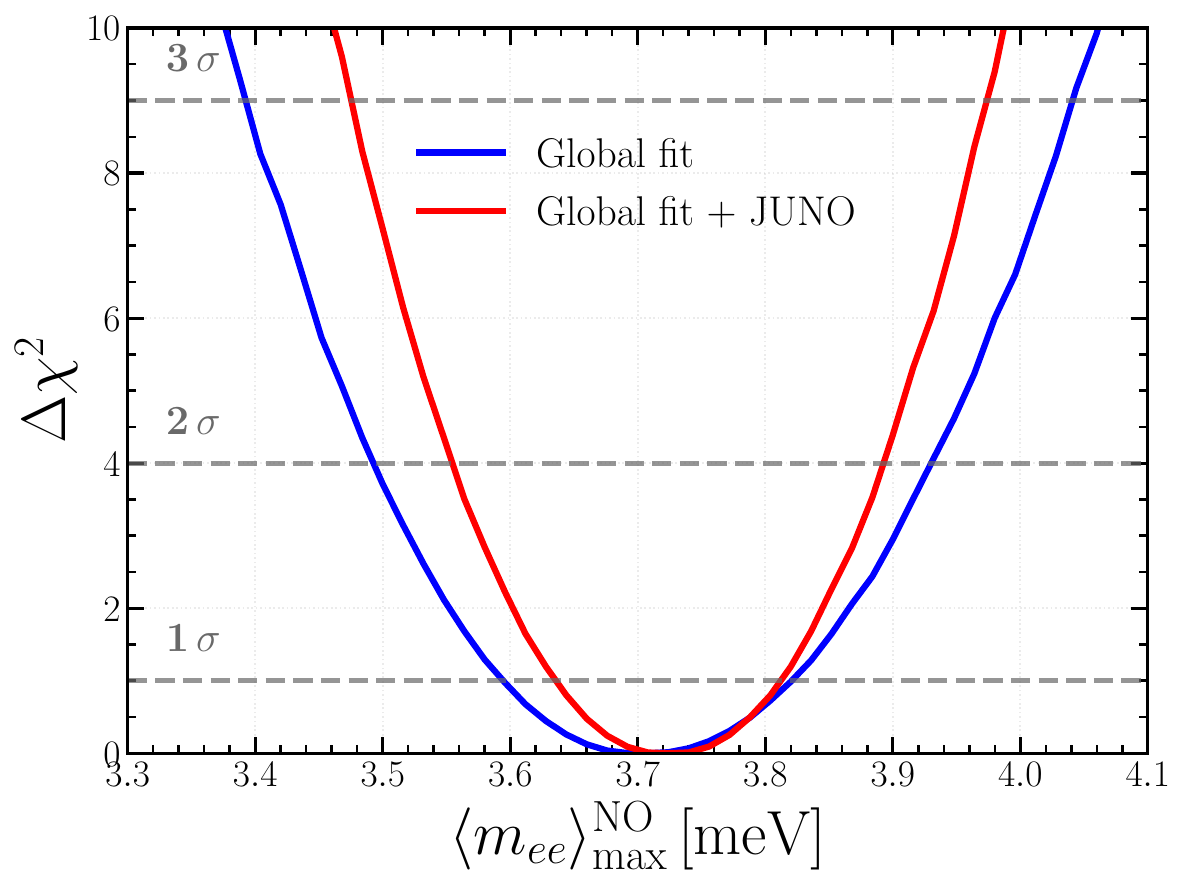}
\caption{One-dimensional $\Delta \chi^2$ for
the $0 \nu 2 \beta$ effective mas limits
$\langle m_{ee} \rangle^\text{IO}_\text{min}$ (left),
$\langle m_{ee} \rangle^\text{NO}_\text{min}$ (center), 
and $\langle m_{ee} \rangle^\text{NO}_\text{max}$ (right)
according to the neutrino oscillation
global fit (blue) \cite{NuFIT6} or the combined results with
both global fit and JUNO (red) \cite{JUNO1st}.}
\label{fig:meeimprovement}
\end{figure}

To make the influence of the JUNO first data \cite{JUNO1st} on the $\nuless$
effective mass limits more explicit, \gfig{fig:meeimprovement}
shows the one-dimensional $\Delta\chi^2$ profiles for 
$\mee^\text{IO}_\text{min}$, $\mee^\text{NO}_\text{min}$, and
$\mee^\text{NO}_\text{max}$ with vanishing lightest mass ($m_\text{lightest}\to0$). 
In this limit, the residual uncertainty is contributed by the
oscillation parameters
$\Dma$, $\Dms$, $\theta_s$, and $\theta_r$, as discussed before. 
The gray dashed horizontal lines indicate the $1\,\sigma$, $2\,\sigma$, 
and $3\,\sigma$ confidence levels ($\Delta\chi^2=1$, $4$, and $9$, respectively). 
For comparison, the blue curves represent the NuFIT 6.0
global fit of neutrino oscillation experiments \cite{NuFIT6}
while the red curves show the improvement from the
JUNO first data \cite{JUNO1st}. In all the three panels,
the improvement is quite sizable which clearly shows
the importance of medium baseline reactor experiments.
In all three panels,
the improvement is quite sizable, which clearly demonstrates
the importance of medium baseline reactor experiments.
Quantitatively, from the narrowing of the $\Delta\chi^2$ profiles,
one can estimate that JUNO has reduced the uncertainties 
on these effective mass limits by approximately
$22.0\%$ for $\mee^{\rm IO}_{\rm min}$, 
$22.5\%$ for $\mee^{\rm NO}_{\rm min}$, and 
$23.1\%$ for $\mee^{\rm NO}_{\rm max}$.
The corresponding $3\,\sigma$ intervals after including 
the JUNO first data are
$\mee^{\rm IO}_{\rm min}=(16.36,\,21.48)\,$meV, 
$\mee^{\rm NO}_{\rm min}=(1.268,\,1.764)\,$meV, and 
$\mee^{\rm NO}_{\rm max}=(3.484,\,3.964)\,$meV.
Although JUNO does not measure the $\nuless$ decay directly
in the first several years, it plays an important role
in reducing the uncertainty from particle physics side.

\begin{figure}[t!]
\centering
\includegraphics[width=0.7\linewidth , angle=-90]{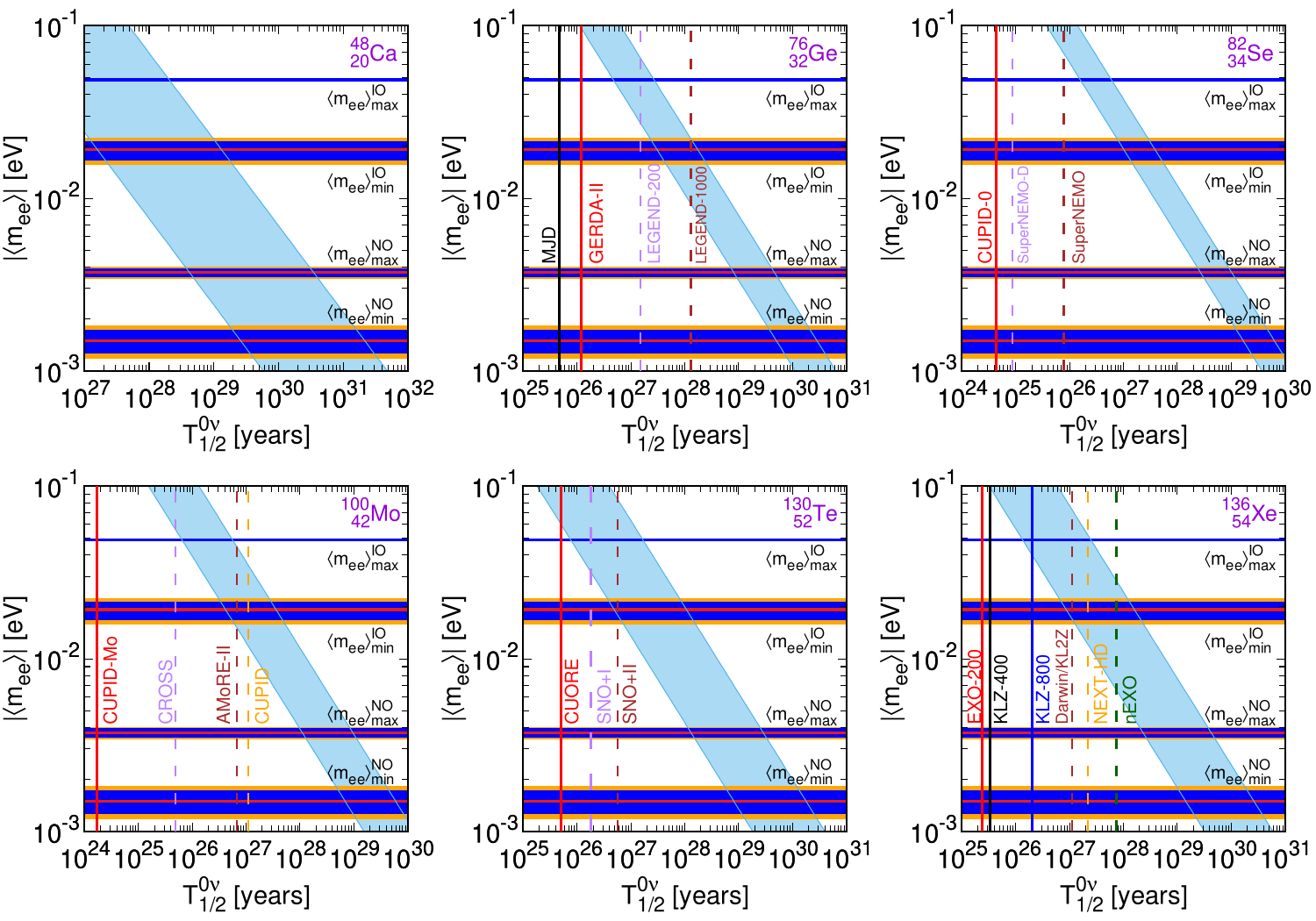}
\caption{
The $0 \nu 2 \beta$ decay half-life time $T^{0 \nu}_{1/2}$
and effective Majorana mass $|\langle m_{ee} \rangle|$
as well as their uncertainties from the nuclear and
particle physics sides. Different panels are for different
isotopes. The light-blue band indicates the nuclear physics 
uncertainty when translating the measured half-life 
$T^{0 \nu}_{1/2}$ into a sensitivity on the $0 \nu 2 \beta$ effective 
mass $|\langle m_{ee} \rangle|$. Current experimental
limits on $T^{0 \nu}_{1/2}$ (90\%\,C.L.) 
are shown as solid vertical lines with the region 
to the left of each solid line is excluded while the
future projections are shown as vertical dashed lines. 
The horizontal bands show the effective mass limits for the vanishing
lightest neutrino mass ($m_\text{lightest}\to0$): 
$\langle m_{ee} \rangle^{\rm IO}_{\rm min}$,
$\langle m_{ee} \rangle^{\rm NO}_{\rm min}$, 
and $\langle m_{ee} \rangle^{\rm NO}_{\rm max}$, with central values 
indicated by the red solid lines. For comparison,
the orange band shows the $3\,\sigma$ ranges
from the current global fit while the blue band 
is for the combined result of global fit plus JUNO.
}
\label{fig:0vbb_halflife}
\end{figure}

Translating the measurement of the half-life
time $T^{0 \nu}_{1/2}$ to the effective mass $\mee$
suffers from the uncertainty in the nuclear
matrix elements (NME) $M^{0\nu}$ in \geqn{eq:Thalf}.
These nuclear physics uncertainties surpass those
from neutrino parameter determinations.
Following the framework established in \cite{GE1},
the Fig.\,3 therein is updated as \gfig{fig:0vbb_halflife}
with the current NME calculations extracted from
\cite{Agostini:2022zub}. \gfig{fig:0vbb_halflife} illustrates
the current experimental limits on $\mee$ versus $\Thf$
for six isotopes: $^{48}$Ca, $^{76}$Ge, $^{82}$Se,
$^{100}$Mo, $^{130}$Te, and $^{136}$Xe. The light blue
bands indicate the nuclear uncertainties using the
combined NME range of available calculations for
each isotope (see Table I of \cite{Agostini:2022zub}).
Even with fixed value for the half-life time $T^{0 \nu}_{1/2}$,
there is no single fixed value of the effective mass
$\mee$ to be extracted.

The current experimental sensitivities on the half-life
time $T^{0 \nu}_{1/2}$ at 90\% C.L. are represented by
the solid vertical lines in \gfig{fig:0vbb_halflife}.
Regions to the left-hand side of the vertical line
have been excluded. Among various measurements, the most stringent
constraints come from KamLAND-Zen ($^{136}$Xe) with
$\Thf > 2.0 \times 10^{26}$\,yr
\cite{KamLAND-Zen:2024eml,KamLAND-Zen:2022tow,KamLANDZen:2016pfg}
and GERDA ($^{76}$Ge) with $\Thf > 1.2 \times 10^{26}$\,yr
\cite{GERDA:2017foj}. Other competitive bounds include
MAJORANA DEMONSTRATOR ($^{76}$Ge, $\Thf > 3.9 \times 10^{25}$\,yr)
\cite{Majorana:2022udl}, EXO-200 ($^{136}$Xe,
$\Thf > 2.4 \times 10^{25}$\,yr) \cite{EXO200:2019rkq},  PandaX-4T ($^{136}$Xe,
$\Thf > 2.1 \times 10^{24}$\,yr) \cite{PandaX:2024fed},
CUORE ($^{130}$Te, $\Thf > 4.8 \times 10^{25}$\,yr)
\cite{CUORE:2022xwt}, CUPID-Mo ($^{100}$Mo,
$\Thf > 9.7 \times 10^{24}$\,yr) \cite{CUPID:2022okx},
and CUPID-0 ($^{82}$Se, $\Thf > 4.4 \times 10^{24}$\,yr)
\cite{CUPID:2022okx}.

Comparing these experimental limits
with the horizontal bands in \gfig{fig:0vbb_halflife}, 
we observe that the current experiments are beginning to
probe the IO parameter space, approaching the lower edge of 
$\mee^{\rm IO}_{\rm min}$. However, they remain several orders 
of magnitude away from the NO region that extends from 
$\mee^{\rm NO}_{\rm min}$ to $\mee^{\rm NO}_{\rm max}$, 
in the limit of vanishing lightest neutrino mass 
($m_\text{lightest}\to 0$). The orange and blue bands 
represent the oscillation parameter uncertainties for each 
of these predictions, showing how the JUNO measurement 
(blue band) narrows the allowed ranges compared 
to current global fit alone (orange band).
For NO, the situation is more challenging. The maximum 
allowed value $\mee^{\rm NO}_{\rm max}$ might be accessible 
to the next-generation experiments while the minimum value 
$\mee^{\rm NO}_{\rm min}$ lies at much lower masses, 
requiring significantly better sensitivities than currently 
projected.

The effort for next-generation experiments promises
a qualitative leap in sensitivity, represented by the
dashed vertical lines in \gfig{fig:0vbb_halflife},
with the flagship experiments nEXO ($^{136}$Xe)
\cite{nEXO1,nEXO2} and LEGEND-1000 ($^{76}$Ge) \cite{LEGEND}.
Both are projected to reach an unprecedented half-life
sensitivity of $\Thf \sim 10^{28}$\,yr.
As illustrated by the intersection of their dashed
vertical lines with the light blue NME uncertainty band,
the corresponding $\mee$ sensitivity ranges are
$(6 \sim 27)$\,meV for nEXO and $(9 \sim 21)$\,meV for LEGEND-1000.
In addition, the CUPID experiment \cite{CUPID1,CUPID2}
targeting $^{100}$Mo and $^{82}$Se aims for
$\Thf \sim 10^{27}$\,yr which can translate to
$\mee = (12 \sim 34)$\,meV. We also include in
\gfig{fig:0vbb_halflife} the projected sensitivities
of DARWIN \cite{DARWIN2}, KamLAND-Zen \cite{KamLANDZen6},
NEXT-HD \cite{NEXT2}, SNO+ \cite{SNOplus2},
AMoRE-II \cite{AMoRE}, SuperNEMO \cite{SuperNEMO1},
and LEGEND-200 \cite{LEGEND}.

While these next-generation experiments represent impressive
technical achievements, their projected sensitivities would only 
begin to probe the IO region near $\mee^{\rm IO}_{\rm min}$,
assuming favorable NME calculations. Definitive exclusion of  
IO would require reduction of the NME uncertainties or 
further experimental improvements beyond the current projections.
For the NO case, these upcoming experiments remain far from
accessing most of the allowed parameter space. Even with 
optimistic assumptions, nEXO and LEGEND-1000 would only marginally
reach $\mee^{\rm NO}_{\rm max}$, while the region near 
$\mee^{\rm NO}_{\rm min}$ would remain inaccessible. 
This highlights the critical importance of reducing the 
oscillation parameter uncertainties though future measurements.
The combination of improved $\nuless$ sensitivity and 
precise oscillation data with reactor neutrino experiments,
particularly from JUNO, will be essential for meaningful
constraints on the effective mass and the neutrino mass ordering.

\section{JUNO Impact on Majorana Triangle and Majorana CP Phases}
\label{sec:majo_triang}

The particle physics contribution to the $\nuless$
decay effective mass $\mee$ is not just the two mixing
angles ($\theta_r$ and $\theta_s$) and the two mass
squared differences. There are two Majorana CP phases
$\dMa$ and $\dMc$ which also come from the particle
physics side. Not to say the lightest neutrino mass or the
absolute mass scale. However, the neutrino oscillation
experiments cannot measure neither of them. While the
neutrino mass scale can be probed by the beta decay
spectrum experiment KATRIN and the cosmological
observations as summarized in \gsec{sec:massScale},
the only chance for measuring the two
Majorana CP phases comes from the $\nuless$ decay
itself. We will investigate how the improved oscillation
parameter measurements affect the prospect of measuring
the two Majorana CP phases through the Majorana triangle
in \gsec{sec:MajoranaCP}.

\subsection{Cosmological Measurement of Neutrino Mass Scale and Preference for NO}
\label{sec:massScale}

\begin{figure}[t]
\centering
\includegraphics[width=0.48\linewidth]{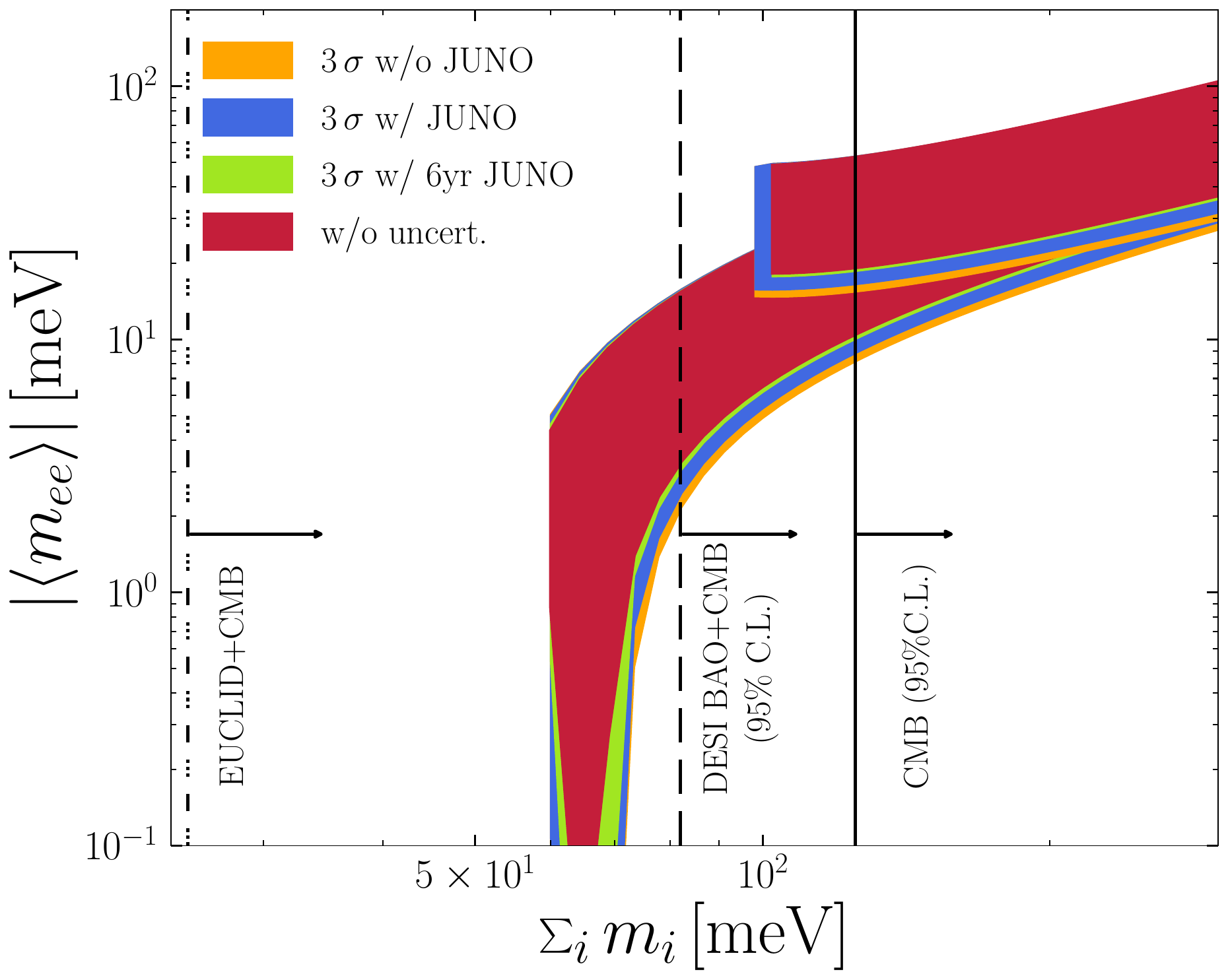}
\hfill
\includegraphics[width=0.48\linewidth]{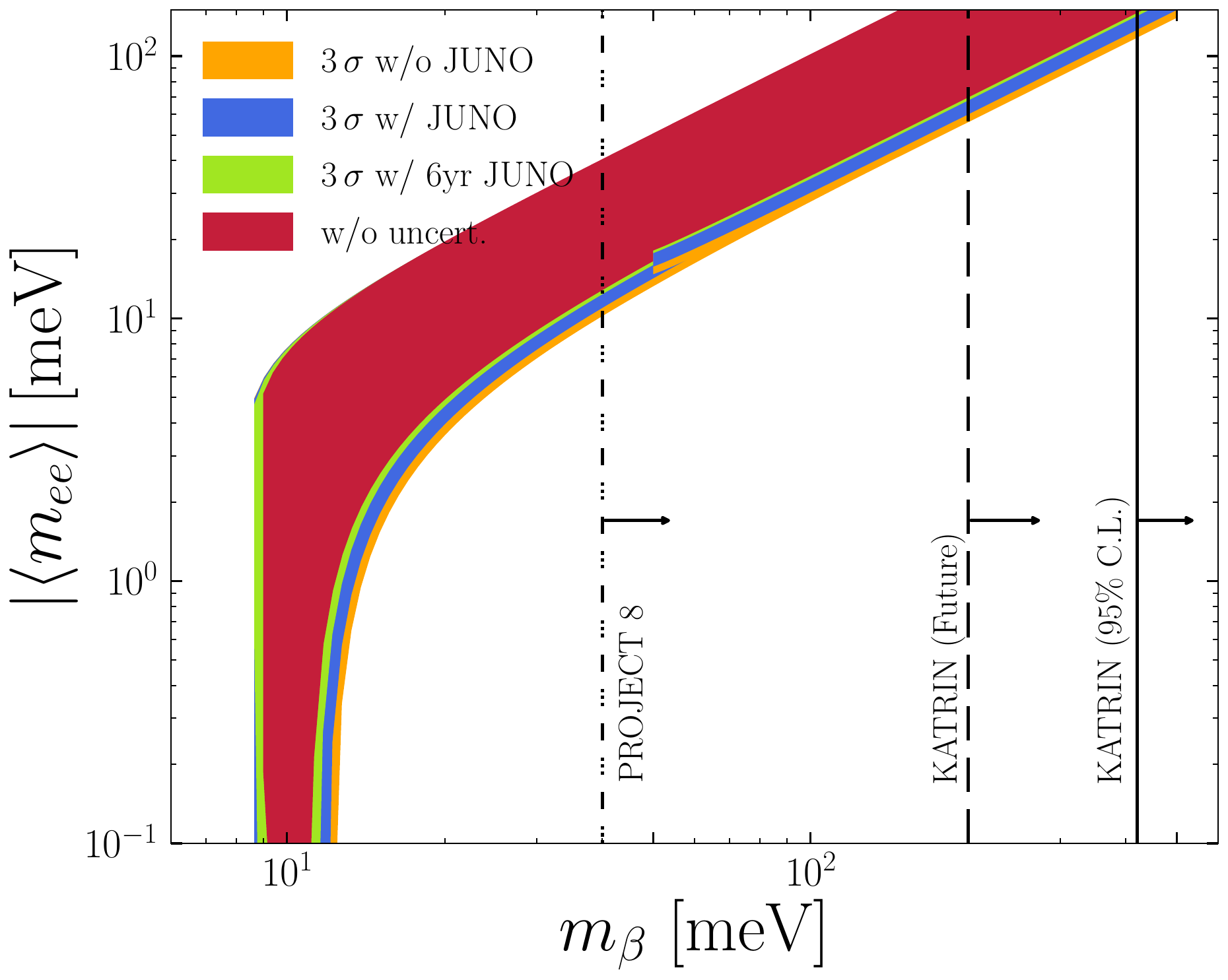}
\caption{{\bf (Left)} Cosmological constraints on the
neutrino mass sum $\sum_i m_i$ at 95\%\,C.L. with vertical line
for the exiting observations from CMB (solid) and
DESI BAO (dashed). For comparison, the projected sensitivities
at the future EUCLID is shown as dash-dotted line.
{\bf (Right)} The current KATRIN (solid) and its
design (dashed) sensitivities on the beta decay
effective mass $m_\beta$. For comparison, the expected
PROJECT 8 sensitivity is also shown as a vertical dash-dotted line.
The uncertainties of the effective Majorana mass $\langle m_{ee} \rangle$
from the particle physics side are also shown according
to the the color scheme of Fig.\,1.}
\label{fig:cosmoMass}
\end{figure}

The cosmological constraints on the neutrino mass sum $\ms$
provide complementary information to laboratory searches.
Especially, the DESI BAO measurements with CMB data gives
the most stringent current bound, $\ms < 0.082$ eV at $95\%$\,C.L.
\cite{DESI2024b,DESI2024}.
For reference, the combination of CMB data from Planck,
the non-DESI Baryon Acoustic Oscillations (BAO) observation,
Cosmic Chronometers (CC), and Pantheon, yields $\ms < 0.12$\,eV
at $95\%$\,C.L. \cite{Planck:2018vyg}.
These bounds are indicated in the left panel of
\gfig{fig:cosmoMass} with vertical lines. The current
cosmological observations have already went beyond the
IO regime to probe its NO counterpart. The future cosmological
surveys, such as EUCLID, are expected to reach the sensitivity
of $\ms < 0.056$\,eV \cite{EUCLID2020,EUCLID2022} which
would directly test the global lower limit of the neutrino
mass sum.

However, the cosmological measurement highly depends on
the data set adopted in the analysis and the theoretical
assumptions or systematics. When systematic uncertainties
in the determination of cosmological parameters such as
$\sigma_8$ and $S_8$ are properly accounted for
\cite{Vagnozzi:2017ovm,Gariazzo:2022ahe,DiValentino:2020vvd}, or when
extended cosmological models such as the time-varying
dark energy \cite{Lorenz:2017fgo}, dark matter-dark radiation
interactions \cite{Archidiacono:2019wdp}, massive neutrino
self-interactions \cite{RoyChoudhury:2019hls,Forastieri:2019cuf}, or other
beyond $\Lambda$CDM scenarios \cite{Garny:2020rom}
are considered, the constraints can be relaxed to
$\ms \lesssim (0.1\sim0.15)$\,eV, where the IO regime is still allowed.
Besides the conventional CMB and matter power spectrum with
linear dependence $\sum_i m_i$,
the cosmic gravitational focusing with fourth power dependence
$\sum_i m^4_i$ can provide an independent cosmological
measurement of the neutrino masses \cite{Ge:2023nnh}
and mass ordering \cite{Ge:2024kac}.

The kinematic constraint $m_\beta < 0.45$\,eV at
$90\%$\,C.L. \cite{KATRIN2022,KATRIN2024} on the beta
decay effective mass $m_\beta \equiv \sqrt{\sum_i |U_{ei}|^2m_i^2}$
from the KATRIN experiment provides a direct and model-independent
measurement of the neutrino mass scale. Although the beta
decay constraint is not as strong as the cosmological
observation, we also show them
in the right panel of \gfig{fig:cosmoMass} for comparison.
The current laboratory measurements probe the
cosmologically allowed mass range, but still remain about an
order of magnitude above the sensitivity required to
test the IO scenario. The KATRIN experiment is expected
to reach its design sensitivity $m_\beta \sim 0.2$\,eV
at $90\%$ C.L. after several years of data taking
\cite{KATRINFuture}. The next-generation experiments
such as PROJECT 8 are projected to reach the sensitivity
of $m_\beta \sim 0.04$\,eV \cite{Project8,Project8Roadmap},
which would probe the majority of the IO regime.

\begin{figure}[t!]
\centering
\includegraphics[width=0.32\linewidth]{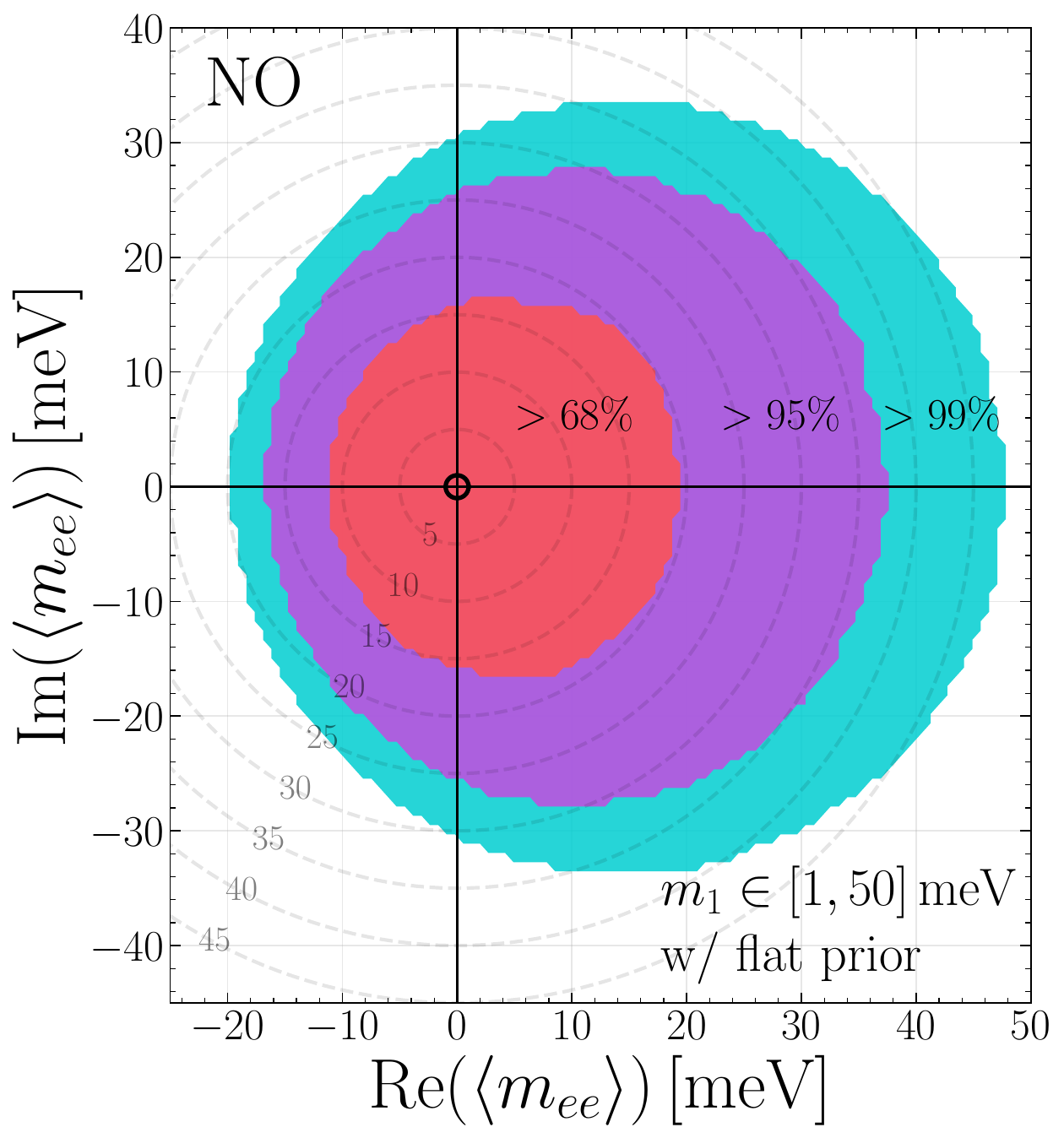}
\includegraphics[width=0.32\linewidth]{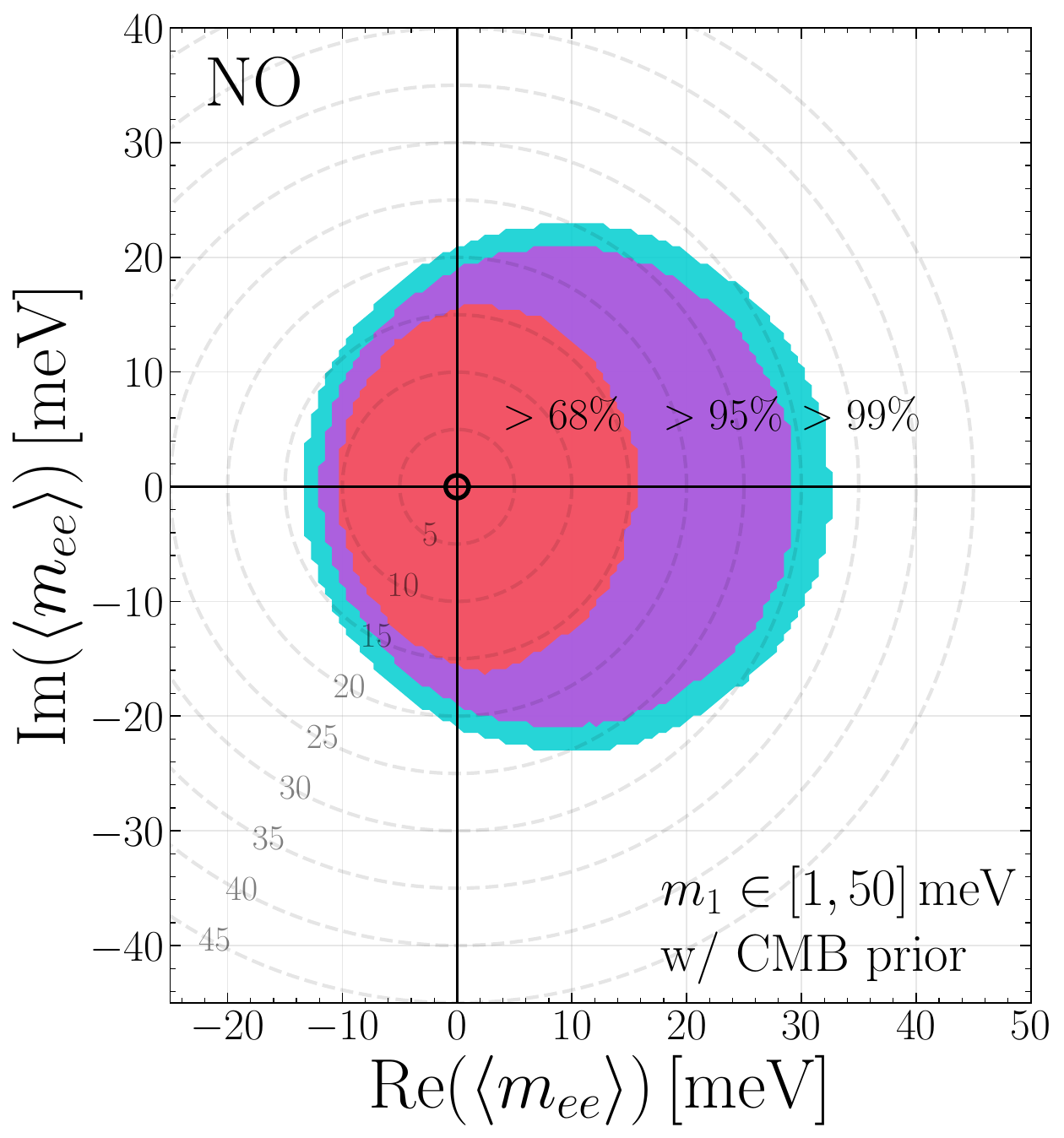}
\includegraphics[width=0.32\linewidth]{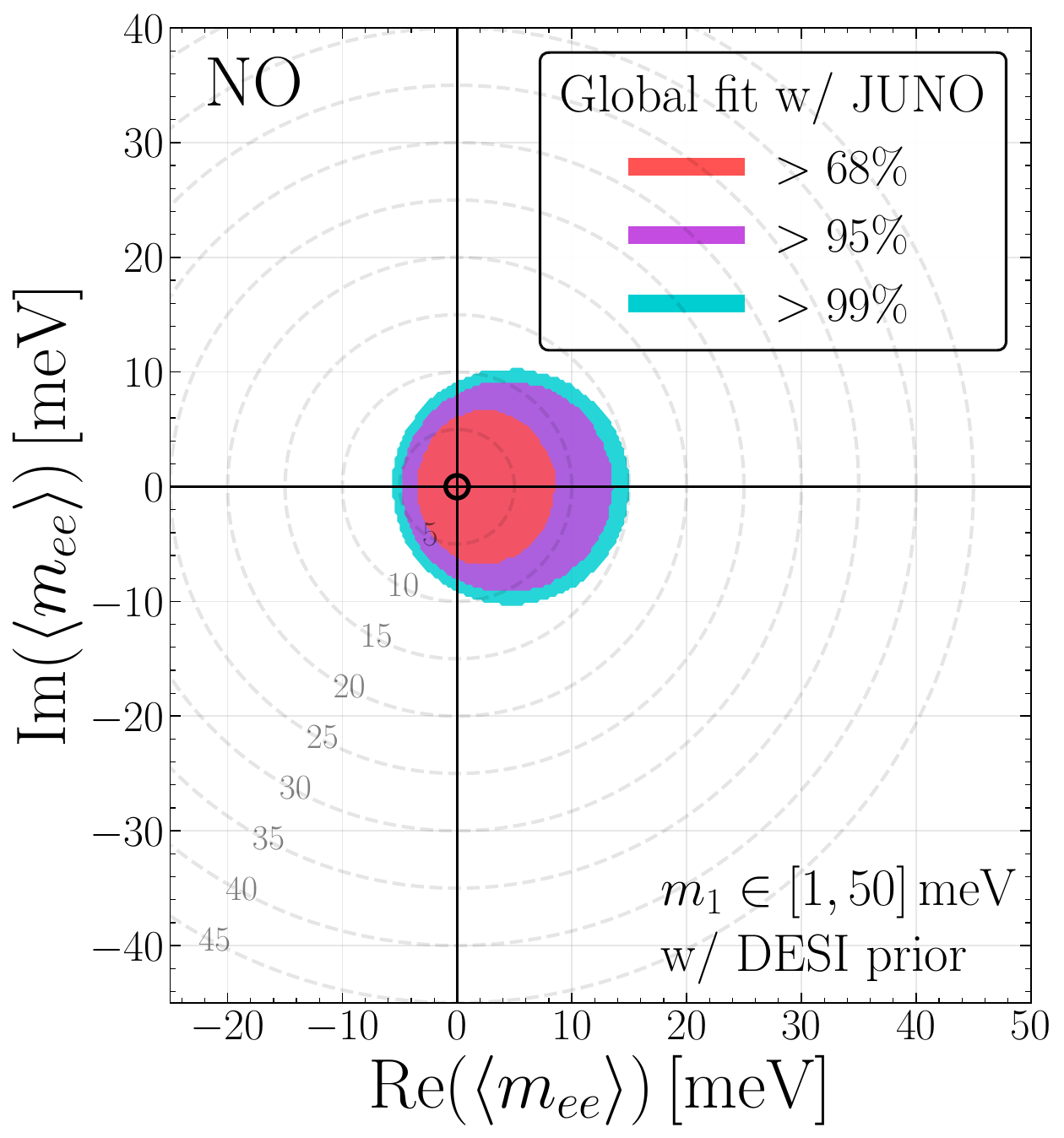}
\caption{Allowed regions in the complex plane of
the effective mass $\langle m_{ee} \rangle$ (real vs. imaginary parts) for NO. 
Three scenarios are presented: 
{\bf (Left)} $m_{\text{lightest}} \in [1,50]$\,meV with a flat prior
and no cosmological bounds on $\sum_i m_i$; 
{\bf (Center)} $m_{\text{lightest}} \in [1,50]$\,meV with 
a prior from the pre-DESI CMB cosmological bounds on $\sum_i m_i$; and 
{\bf (Right)} $m_{\text{lightest}} \in [1,50]$\,meV with 
a prior from DESI BAO~+~CMB cosmological bounds on $\sum_i m_i$. 
The colored contours represent different confidence 
levels: red (68\%\,C.L.), purple (95\%\,C.L.), and 
cyan ($99\%$\,C.L.). The black circle around the origin
indicates the funnel region with $|\langle m_{ee} \rangle| < 1$\,meV.}
\label{fig:cosmoMee}
\end{figure}

As summarized in \gsec{sec:theory0vbb} and in particular
\geqn{eq:triangle}, the effective
mass $\mee$ can be viewed as a vector sum of three
contributions in the complex plane. 
With our phase convention $\dMb = 0$, the vector $\vLb$ lies along the real axis while $\vLa$ and $\vLc$ rotate 
with the two physical Majorana CP phases $\dMa$ and $\dMc$.
This geometrical configuration is finally determined by the 
allowed ranges for $|U_{ei}|^2$ from oscillation experiments,
the relative magnitudes of the triangle sides
$L_i = m_i|U_{ei}|^2$, and the absolute mass scale
$m_{\text{lightest}}$.

The behavior of $\mee$ in the complex plane reveals
the fundamental difference between the two mass orderings. 
For NO, the mass hierarchy $m_1 < m_2 \ll m_3$  allows
these three vectors to form a closed triangle that can
reach complete cancellation, $\mee = 0$, 
for specific combinations of the two Majorana CP phases
\cite{Vissani:1999tu,Xing:2014yka,GE2}.
In contrast, for IO with $m_3 \ll m_1 < m_2$, 
the relative magnitudes of the vector sides prevent
such closure, making $\mee = 0$ geometrically impossible 
regardless of the Majorana phases. Such fundamental
difference manifests topologically in the complex
$\mee$ plane for NO.

\gfig{fig:cosmoMee} shows the allowed regions for the
effective mass $\mee$ with the horizontal and vertical
axes denoting its real and imaginary parts, respectively.
As the Majorana phases $\dMa$ and $\dMc$ vary over
their full range $[0, 2\pi]$, the allowed values of
$\mee$ trace out a characteristic circle region,
which can be understood with the geometrical picture.
The real $L_2$ first shifts to the central point to
the right. Based on this, the vector $\vLa$ with length
$L_1 = m_1 c^2_s c^2_r$ rotates around the shifted
epicenter and the third vector $\vLc$ with shorter
length $L_3 = m_3 s^2_r$ further scanning the whole
plane. Vertically, the allowed regions are symmetric
around the horizontal axis which is a manifestation of
the fact that the imaginary part can flip sign when
the two Majorana CP phases change their signs simultaneously.

Across the three panels of \gfig{fig:cosmoMee},
the size of the circles differ considerably. The major
factor is the allowed range of the neutrino mass
scale. For comparison, the left panel
shows the results obtained with even distribution (flat prior)
of the lightest mass in the range of
$m_1 \in [1, 50]$\,meV. Once the cosmological
constraints are imposed in the center and right
panels, the allowed circle regions shrink.
The effect is not that significant for the
case with only pre-DESI constraints since
the limit $\sum_i m_i < 0.12$\,eV 95\%\,C.L. \cite{Planck:2018vyg}
translates to $m_1 \lesssim 30$\,meV which
is roughly the size of circle. To make the 
connection more direct, the corresponding
upper limit for the NO effective mass in
\geqn{eq:mee_IO} with a nonzero $m_1$ is,
\begin{align}
  \mee^{\rm NO}_{\rm max}(m_1)
& = 
  m_1 c^2_s c^2_r
+ \sqrt{\Dms + m_1^2} c^2_r s^2_s
+ \sqrt{\Dma +m_1^2} s^2_r
\approx
  31\,\mbox{meV}.
\end{align}
However, once the DESI constraint
$\sum m_i < 0.082$\,eV at 95\%\,C.L.\cite{DESI2024b,DESI2024}
is imposed, the allowed circle significantly
shrinks to $\mee^{\rm NO}_{\rm max}(m_1) \approx 16$\,meV
with $m_1 \lesssim 14$\,meV.

A very important feature is that the vanishing effective
mass scenario, $\mee = 0$ which corresponds to the origin
of the plots, is always allowed by the current constraints.
This because the funnel or throat region shown in the left
panel of \gfig{fig:0vbb_comparison} is covered by the
sampling regions, $m_1 \in [1, 50]$\,meV. Even the DESI result with
$m_1 \lesssim 14$\,meV still covers this funnel region
as clearly indicated by the dashed line in the left panel
of \gfig{fig:cosmoMass}. Most importantly, the funnel
region $|\mee| < 1$\,meV indicated by the black circle
around the origin is actually in the most probable range
(red band).
With more stringent bound from DESI, the relative size
of the funnel region with respect to the whole circle actually
increases. It indicates that the chance for the $\nuless$
decay effective mass $\mee$ to fall into the funnel
region is actually increasing.

\subsection{Determination of Two Majorana CP Phases}
\label{sec:MajoranaCP}

\begin{figure}[t!]
\centering
\includegraphics[width=0.8\linewidth]{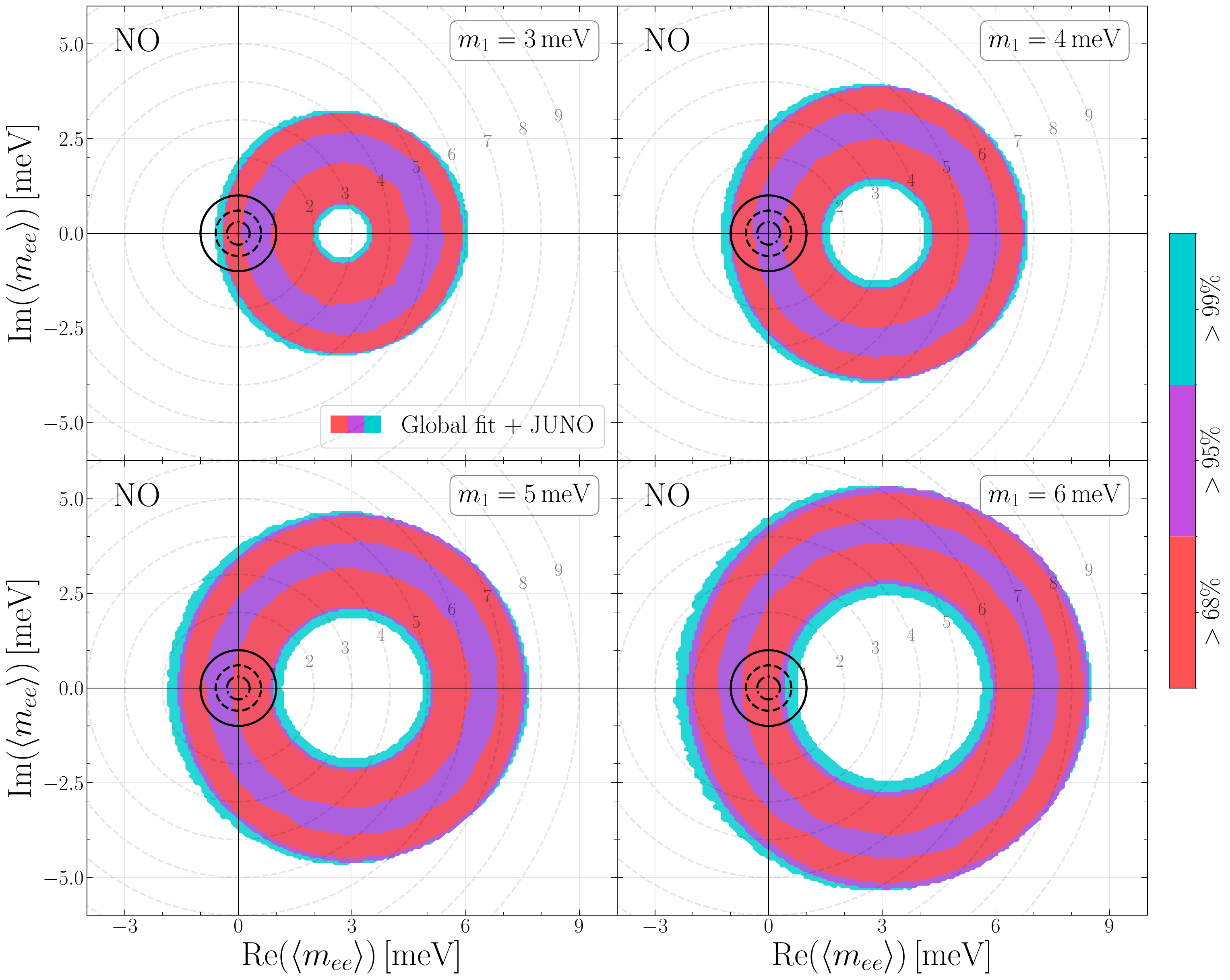}
\caption{Allowed regions in the complex plane of $\langle m_{ee} \rangle$ 
(real vs. imaginary parts) for NO with the lightest mass
$m_1 = 3$, 4, 5, 6\,meV. The 
filled color regions represent the 68\% (red),
95\% (purple), and 99\% (cyan) confidence levels.
The black circles indicate the funnel region with
$|\langle m_{ee}\rangle| < 1$\,meV (solid),
0.6\,meV (dashed), and 0.3\,meV (dash-dotted).}
\label{fig:mee_plane}
\end{figure}

As shown above, the scenario of a vanishing effective mass
$\mee$ or being bound from above has visible chance.
To focus on such scenarios, \gfig{fig:mee_plane} shows
the allowed regions in the complex $\mee$ plane for NO
with four representative values of the lightest mass, 
$m_{\text{lightest}} = (3$, 4, 5, 6)\,meV in different
panels. The figure clearly illustrates the characteristic
circle topology for NO that passes through the origin. 
For $m_{\text{lightest}}$ in the narrow window of $(3 \sim 6)$\,meV,
the 68\%\,C.L. region (red contours) covers the origin. 
In other words, the three vectors can arrange themselves
into a closed triangle with complete cancellation at
specific combinations of $\dMa$ and $\dMc$ to satisfy
the geometrical closure conditions.
A major difference from \gfig{fig:cosmoMee} is that
now the circle is not fully filled but has a black
region in the middle. This is because with fixed
$m_1$, the three vector lengthes $L_i$ cannot adjust
freely, especially for $L_2$ whose length determines
the thickness of the doughnut.

If the cancellation among the three complex vectors
$\overrightarrow{L_i}$ really happens, a {\it Majorana triangle}
can form. Then it is possible to determine the
two Majorana CP phases simultaneously as function of
the three vector lengths \cite{GE2},
\begin{eqnarray}    
  \cos \dMa
=
- \frac {L_1^2 + L_2^2 - L_3^2}
        {2L_1L_2},
\label{eq:cancel_alpha}
\qquad
  \cos \dMc
=
  \frac {L_1^2 - L_2^2 - L_3^2}
        {2L_2L_3}.
\label{eq:cancel_beta}
\end{eqnarray}
Note that only a single degree of freedom, as a combination
of the two Majorana CP phases, can be determined for
a nonzero effective mass. However, a vanishing $\mee$
requires both its real and imaginary parts to vanish.
In other words, two independent conditions can appear
in the vanishing limit to simultaneously constraint
the two Majorana CP phases. This is a very rare example
that one extra constraint can appear. Even if there
is experimental uncertainty and hence only an upper
bound on the effective mass can be extracted,
$|\mee| < c$, it can transfer to both the imaginary
and real parts, $|\mbox{Re}(\mee)| < c$ and
$|\mbox{Im}(\mee)| < c$. The funnel region has been
shown with black circles with $c = 1$\,meV (solid),
0.6\,meV (dashed), and 0.3\,meV (dash-dotted) in
\gfig{fig:mee_plane}.

A paradox here is that if
the $\nuless$ decay is observed, one physical degree
of freedom cannot be observed. But if the effective
mass $\mee$ vanishes to make the $\nuless$ decay
invisible, both physical Majorana CP phases can be
simultaneously determined. The Majorana nature can
be probed with other ways, such as the direct detection
of cosmic neutrino backgrounds with tritium
\cite{Ringwald:2009bg,Vogel:2015vfa,Shergold:2021evs}
or the atomic radiative emission of neutrino pairs
\cite{Dinh:2012qb,Fukumi:2012rn}.

Besides \cite{GE2}, determining the Majorana CP phases
according to \geqn{eq:cancel_beta} as cosine functions of
the Majorana triangle sides has also been explored
in \cite{Xing:2015zha} and \cite{Cao:2019hli}. Both \cite{GE2}
and \cite{Cao:2019hli} has projected the uncertainty
reduction by the JUNO experiment. In our current study,
we would use the first data release from JUNO \cite{JUNO1st} to make
realistic illustration of the improvement that the JUNO
experiment can already provide.

\begin{figure}[t!]
\centering
\includegraphics[width=0.6\linewidth]{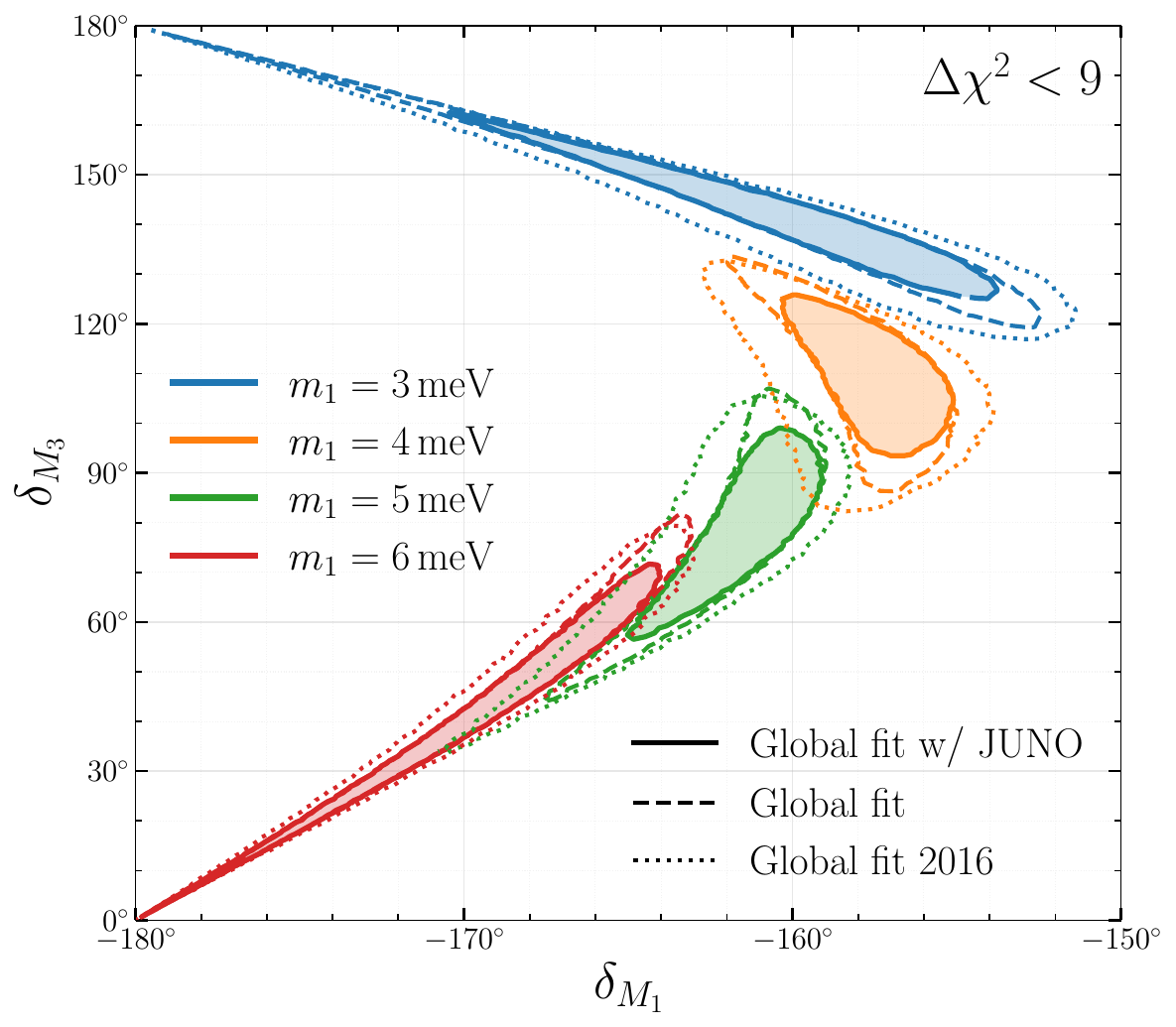}
\caption{The two Majorana CP phases $(\delta_{\rm M1}, \delta_{\rm M3})$
determined by the Majorana triangle ($\langle m_{ee} \rangle = 0$)
for NO with $m_1 = 3$\,meV (blue), 4\,meV (orange),
5\,meV (green), and 6\,meV (red). Three uncertainty
schemes of the oscillation parameters with
$\Delta \chi^2 < 9$ are shown. The dotted contours
represent results from the Fig.\,7 of \cite{GE2},
dashed contours show the current NuFIT 6.0 bounds,
while the thick solid contours incorporate the
latest JUNO measurement.}
\label{fig:majorana_phases}
\end{figure}

\gfig{fig:majorana_phases} illustrates this possibility
by showing the two Majorana CP phases $\dMa$ and $\dMc$
determined by the Majorana triangle ($\mee = 0$) for NO
with different values of the lightest neutrino mass
$m_1 = 3$, 4, 5, 6\,meV. Each colored region represents
one allowed solution: blue for 3\,meV, orange for
4\,meV, green for 5\,meV, and red for 6\,meV.
These solutions are obtained by varying the oscillation
parameters within $\Delta\chi^2 < 9$ and solving
\geqn{eq:cancel_alpha} for each parameter combination. 
Three uncertainty schemes are displayed: the
previous uncertainties used in the Fig.\,7 of
\cite{GE2} (dotted contours), current NuFIT 6.0
bounds \cite{NuFIT6} (dashed contours), and
the combination of NuFIT with the JUNO
first measurement \cite{JUNO1st} (thick solid contours).
The progressive narrowing of these allowed
regions, particularly evident when comparing
the dotted with the solid contours, directly
demonstrates how the reduced uncertainties of
oscillation parameters translate into more
precise predictions for the Majorana phases
solutions.

The comparison in \gfig{fig:majorana_phases} clearly demonstrates
how the reactor experiments can help mitigating the uncertainty
in $\nuless$ decay. The factor of $\sqrt{2}$ reduction in the
$\theta_s$ uncertainty at JUNO narrows the allowed region of
$\dMa$ and $\dMc$ for each value of $m_1$. This reduction is
particularly pronounced because $\theta_s$ directly affects
the relative magnitudes of $L_1$ and $L_2$ through the prefactors
$c_s^2$ and $s_s^2$. The narrower bands in \gfig{fig:majorana_phases}
translate directly into more precise predictions. If future
$0\nu\beta\beta$ experiments do not observe a signal despite
reaching sensitivities $|\mee| \lesssim 1$\,meV,
JUNO's improved oscillation parameters would enable a much
sharper determination of the Majorana CP phases under the
assumption of complete cancellation. The effect of
the $\mee$ experimental uncertainties on the Majorana CP
phase determination
can be found in \cite{GE2} and \cite{Huang:2020mkz}.

\section{Conclusion}
\label{sec:conclusion}

In this paper, we have investigated the impact of the JUNO
first data release on the $\nuless$ decay effective mass and the 
determination of Majorana CP phases. The improved precision on
the oscillation parameters, particularly almost a factor of
$25.9\%$ reduction in the uncertainty of $\sin^2\theta_s$
and $37.3\%$ reduction in $\Delta m^2_s$ 
at $3\,\sigma$ range, directly translates into 
22\% (23\%) reduction in the uncertainty of the
minimum effective Majorana mass for IO (NO). At the $3\sigma$ 
confidence level, we find $\mee^{\rm IO}_{\rm min} = (16.36 \sim 21.48)$\,meV, 
with the narrowed range providing a more precise target for the next-generation 
$0\nu\beta\beta$ experiments. This improvement is crucial for the experimental 
design of facilities such as nEXO, LEGEND-1000, and CUPID, which aim to probe 
the IO parameter space. The sharpened lower bound means that a null 
result from these experiments, combined with JUNO's precision measurements, would 
provide increasingly strong evidence against the IO scenario.
Nevertheless, the JUNO experiment can provide direct
measurement of the neutrino mass ordering.

Beyond the bounds on $\mee$, the improved measurement of
oscillation parameters at JUNO can also significantly enhance 
our ability to determine the Majorana CP phases $\dMa$ and $\dMc$ 
in the scenario of complete cancellation ($\mee =0$) for NO.
For the narrow mass window $m_1 = 3 \sim 6$\,meV where a 
vanishing effective mass can exist, the allowed regions in
the $(\dMa, \dMc)$ plane are substantially reduced from
to previous analyses. Although simultaneous determination
of the two Majorana CP phases is possible if future 
$\nuless$ experiments do not observe a signal when
reaching the sensitivity of $|\mee| \lesssim 1$\,meV,
it still relies on an experiment such as JUNO to reduce
uncertainties from the particle physics side.

Looking ahead, the JUNO full dataset with several years of operation is expected to 
achieve sub-percent precision on $\sin^2\theta_s$, which would further narrow the 
theoretical predictions for $\mee$ in both mass orderings.
This improvement will 
be especially valuable when combined with the next-generation 
$\nuless$ experiments as well as the absolute neutrino mass
measurement from cosmological observations (such as DESI, EUCLID, 
and future CMB experiments) and beta decay experiments (such as PROJECT 8). 
The synergy of these three complementary measurements — the precision neutrino
oscillation, the $\nuless$ decay, and the absolute mass scale determinations —
forms an essential triad for fully characterizing the nature of neutrino
masses. The results presented in this work demonstrate that the precision 
era of neutrino physics has truly begun, with JUNO playing a central role in 
shaping the landscape of $\nuless$ decay phenomenology for the coming decade.

\section*{Acknowledgements}
The authors would like to thank Yue Meng for useful discussions.
The authors are supported by the National Natural Science
Foundation of China (12425506, 12375101, 12090060, and 12090064)
and the SJTU Double First
Class start-up fund (WF220442604).
CFK is supported by IBS under the project code IBS-R018-D1.
SFG is also an affiliate member of Kavli IPMU, University of Tokyo.

\bibliography{references}
\bibliographystyle{utphysGe}

\end{document}